\documentclass[sigconf,nonacm,9pt]{acmart}

\newcommand\vldbdoi{XX.XX/XXX.XX}

\newcommand\vldbavailabilityurl{http://vldb.org/pvldb/format_vol14.html}

\usepackage[english]{babel}
\usepackage{caption}
\usepackage{subcaption}
\usepackage{algpseudocode}
\usepackage{lscape}
\usepackage[linesnumbered,ruled,vlined]{algorithm2e}

\usepackage{amsmath}
\usepackage{graphicx}
\usepackage{cleveref}
\usepackage{soul}

\def\HS{\hspace{\fontdimen2\font}}
\definecolor{darkgreen}{rgb}{0.15,0.55,0.15}
\definecolor{darkblue}{rgb}{0.15,0.35,0.7}
\definecolor{blue}{rgb}{0.2, 0.6, 1.0}
\definecolor{darkgreen}{rgb}{0.15,0.55,0.15}
\definecolor{mred}{rgb}{1.0, 0.4, 0.5}
\definecolor{grey}{rgb}{0.5,0.5,0.5}
\definecolor{Purple}{rgb}{.75,0,.85}
\definecolor{light-gray}{gray}{0.95}
\definecolor{mid-gray}{gray}{0.85}
\definecolor{darkred}{rgb}{0.7,0.25,0.25}
\definecolor{rose}{rgb}{1.0, 0.01, 0.24}
\newcommand{\red}[1]{\textcolor{red}{#1}}

\newcommand{\blue}[1]{\textcolor{blue}{#1}}

\def\ojoin{\setbox0=\hbox{$\Join$}%
\rule[0.1ex]{.27em}{.4pt}\llap{\rule[1.3ex]{.27em}{.4pt}}}
\def\leftouterjoin{\mathbin{\ojoin\mkern-5.8mu\Join}}

\newcommand{\revise}[1]{#1}

\newcommand{\eat}[1]{}

\newcommand{\stitle}[1]{\vspace{2pt}\noindent\textbf{#1}}

\newcommand{\thd}[1]{\begin{tabular}[x]{@{}c@{}} #1 \end{tabular}}

\newcommand{\cnt}[0]{\emph{COUNT}\xspace}

\newcommand{\sys}[0]{\texttt{JoinBoost}\xspace}
\newcommand{\lgbm}[0]{\texttt{LightGBM}\xspace}
\newcommand{\xgb}[0]{\texttt{XGBoost}\xspace}

\newcommand{\duckdb}[0]{\texttt{DuckDB}\xspace}
\newcommand{\dbmsx}[0]{\texttt{DBMS-X}\xspace}
\newcommand{\pd}[0]{\texttt{Pandas}\xspace}
\newcommand{\lmfao}[0]{\texttt{LMFAO}\xspace}
\newcommand{\mad}[0]{\texttt{MADLib}\xspace}

\newenvironment{myitemize}{\begin{list}{$\bullet$}{\renewcommand{\leftmargin}{0.15in}}}{\end{list}}

\newtheorem{thm}{Theorem}

\newtheorem{ex}{Example}

\newtheorem{example}[ex]{Example}
\newtheorem{definition}[thm]{Definition}

\author{Zezhou Huang}
\email{zh2408@columbia.edu}
\affiliation{
  \institution{Columbia University}
}
\author{Rathijit Sen}
\email{Rathijit.Sen@microsoft.com}
\affiliation{
  \institution{Microsoft}
}
\author{Jiaxiang Liu}
\email{jl6235@columbia.edu}
\affiliation{
  \institution{Columbia University}
}

\author{Eugene Wu}
\email{ewu@cs.columbia.edu}
\affiliation{
  \institution{DSI, Columbia University}
}

\begin{document}

\title{\sys: Grow Trees Over Normalized Data Using Only SQL}

\begin{abstract}
Although dominant for tabular data, ML libraries that train tree models over normalized databases (e.g., \lgbm, \xgb) require the data to be denormalized  as a single table, materialized, and exported.  This process is not scalable, slow, and poses security risks.  In-DB ML aims to train models within DBMSes to avoid data movement and provide data governance.  Rather than modify a DBMS to support In-DB ML, is it possible to offer competitive tree training performance to specialized ML libraries...with only SQL?

We present \sys, a Python library that rewrites tree training algorithms over normalized databases into pure SQL.  It is portable to any DBMS, offers performance competitive with specialized ML libraries, and scales with the underlying DBMS capabilities.
\sys extends prior work from both algorithmic and systems perspectives. Algorithmically, we support factorized  gradient boosting, by updating the $Y$ variable to the residual in the {\it non-materialized join result}.   Although this view update problem is generally ambiguous, we identify {\it addition-to-multiplication preserving}, the key property of variance semi-ring to support $rmse$, the most widely used criterion.   System-wise, we identify residual updates as a performance bottleneck. Such overhead can be natively minimized on columnar DBMSes by creating a new column of residual values and adding it as a projection. We validate this with two implementations on \duckdb, with no or minimal modifications to its internals for portability. 
Our experiment shows that \sys is $3\times$ ($1.1\times$) faster for random forests (gradient boosting) compared to \lgbm, and over an order of magnitude faster than state-of-the-art In-DB ML systems. Further, \sys scales well beyond \lgbm in terms of the \# features, DB size (TPC-DS SF=1000), and join graph complexity (galaxy schemas).

\end{abstract}

\maketitle



\vspace{-3mm}
\section{Introduction}\label{s:intro}

Tree-based models---ranging from decision trees, random forests, and gradient boosting---are recursive space-partitioning algorithms for classification and regression~\cite{murphy2012machine}.  They have shown exceptional effectiveness for tabular datasets, outperforming popular neural network models~\cite{grinsztajn2022tree}.  In fact,  random forests and gradient boosting were rated the most popular model from 2019 to 2021 on Kaggle~\cite{kaggle_survey}.     In response, the ML community has developed optimized tree-training libraries like  \lgbm~\cite{ke2017lightgbm} and \xgb~\cite{chen2016xgboost}, that offer user-friendly API, superior  single-node performance, and compatibility with distributed frameworks  like Dask~\cite{rocklin2015dask} or Spark~\cite{xin2013shark}. 

In practice, however, almost all tabular datasets are normalized and stored in a DBMS.  Using ML libraries introduces performance, usability, and privacy drawbacks. First, the libraries expect a single (typically CSV) training dataset.  Thus, a developer must denormalize the database into a ``wide table'' $R_\Join$, materialize and export $R_\Join$, and load it into the library.  The join materialization cost is prohibitive for all but the simplest schemas and smallest databases---the $1.2GB$ IMDB dataset (\Cref{fig:cluster}) is well over $1TB$ when fully materialized due to N-to-N relationships.   Second, managing the exported data as well as the separate execution frameworks adds considerable operational complexity and is error-prone~\cite{smelcer1995user,shankar2022operationalizing}.  Third, exporting sensitive data can raise privacy concerns~\cite{PDPA}.

Ideally, we would ``move computation to the data'' and train tree-based models directly within the DBMS.  This would let developers manage the entire data lifecycle---preparation, cleaning, analysis, and modeling---within a single DBMS, and benefit from the DBMSes' excellent security, performance, and scalability.
To be broadly useful, we believe an initial In-DB solution should meet three criteria: (C1) be easily portable to any DBMS by translating ML algorithms into vanilla SQL queries, (C2) offer training performance comparable with the SOTA ML libraries (e.g., \lgbm), and (C3) scale to massive data warehouse sizes and complexity.
Unfortunately, these criteria are often in tension.  
Common wisdom and prior results suggest that training tree models via SQL queries is portable but notoriously slow~\cite{xin2013shark,stonebraker2011architecture,hellerstein2012madlib} as DBMSes do not employ model-specific optimizations. However, specialized optimizations~\cite{ke2017lightgbm,chen2016xgboost}
or GPU acceleration~\cite{asada2022share,hu2021tcudb,paul2021database} achieve competitive performance at the expense of portability to general DBMSes.

To optimize training over normalized schemas common in DBMSes, recent works in factorized ML~\cite{schleich2019layered,khamis2020functional,khamis2018ac,olteanu2016factorized,kobis2017learning} avoid materializing joins by treating ML as semi-ring aggregation queries over the join graph; the aggregations can then be pushed through the joins.
This approach potentially provides DBMSes with an edge over conventional ML libraries that face significant expenses for join materialization and export during training.
Thus, it is intriguing to investigate the potential of training tree models in DBMSes using vanilla SQL, with factorized ML optimizations.

Nevertheless, the current applicability of factorized ML to tree-based models is restricted due to both algorithmic and system-related limitations.
\lmfao~\cite{schleich2019layered} describes a factorized algorithm that is limited to decision trees, and does not support the widely used gradient boosting models.   
It rewrites the tree node split algorithm into a batch of group-by aggregations, and uses multi-query optimization to share work between the aggregations.
Unfortunately, it does not support the residual updates needed for gradient boosting models.
Even for single decision tree training, \lmfao fails to exploit considerable work-sharing between parent-child tree nodes.
Finally, \lmfao lacks portability because its performance is tied to a compilation engine specially designed for its batch-wise query execution.   
In our experiments, despite using an off-the-shelf DBMS (\duckdb~\cite{raasveldt2019duckdb}), we train decision trees $1.9\times$ faster than \lmfao.

To this end, we present \sys,
the first In-DB ML system 
designed for tree-based models (decision trees, random forests, and gradient boosting) that fulfill C1-3. 
\revise{Our main contribution lies in the system design and implementation: we study the feasibility of implementing \sys as a Python library that translates ML algorithms into simple Select-Project-Join-Aggregation (SPJA) queries and employs factorized ML as pure query rewriting. Such design ensures portability (C1) and scalability (C3), as the translated SQL queries can run on any single node or cloud-based DBMS.
Nonetheless, this design poses significant challenges. Algorithmically, previous factorized ML systems do not support gradient boosting. Furthermore, while earlier factorized ML systems have demonstrated good performance, their success is closely linked to specialized execution engines. It's unclear whether implementing factorized ML as pure query rewriting can enable existing DBMSes to deliver performance (C2) competitive to specialized ML libraries.}

To support factorized gradient boosting, each iteration trains a decision tree on the {\it residuals}  of the prior tree, requiring updates to $Y$ in the denormalized relation $R_\Join$;  however, {\it $R_\Join$ is not materialized in factorized ML}, and the view update generally requires $O(R_\Join)$~\cite{kotidis2006updates}. 
\revise{To address this, for snowflake schema, we exploit 1-1 mapping between $R_\Join$  with the fact table $F$, to update $Y$ in $F$ directly.
For the more complex galaxy schema, we found that despite challenges in directly updating $Y$ in $R_\Join$, the tree training only relies on aggregates of $Y$ (e.g., count, sum) that could be updated efficiently. Our technique identifies {\it addition-to-multiplication preserving}, the key property to efficiently update aggregates, by rewriting residual updates over $Y$ in $R_\Join$ into a join with an {\it update relation} $U$.}
However, each boosting iteration generates a new {\it update relation} that may create cycles and doesn't scale with the number of iterations. To address this, we introduce  {\it Clustered Predicate Tree} (CPT) that restricts  the tree splits to attributes within the same cluster.  This enables us to train gradient boosting on IMDB datasets, which was previously prohibitive due to the large size of $R_\Join$ (${>}1TB$).

\revise{To assess the viability of implementing factorized ML with pure SQL, we conduct extensive experiments on \duckdb and a popular commercial DBMS. 
Although columnar DBMS has the potential to compete with specialized ML libraries, we find residual updates as the major bottleneck for gradient boosting in current columnar DBMSes: Residual updates require sequential writes to all values in a column.} However, this process is not efficient in existing DBMSes due to a combination of columnar compression, write-ahead-log (WAL), and concurrency control (CC); these are fundamental to DBMSes but unnecessary for gradient boosting. \revise{Disabling WAL, CC, and compression in existing DBMSes directly is challenging as they are deeply integrated into the codes. To match the performance of \lgbm, update performance must be similar to a parallelized write to an in-memory byte array.}

\revise{To minimize DBMS overheads for residual updates (C2) without compromising portability (C1), we explore the approach to create a new column of residual values and adding it as a projection~\cite{stonebraker2018c} on columnar DBMSes. 
We emulate this on \duckdb in two ways, with no or little modification to its internals.}
First, by leveraging \duckdb \pd API~\cite{relationalapi},  we store the fact table in a \pd dataframe, use \duckdb for joins and aggregations, and update the residual by creating a new \pd dataframe column.  This results in a ${\sim}15\times$ improvement in updates, making gradient boosting competitive without modifying \duckdb. However, one drawback is a ${\sim}1.6\times$ slowdown in aggregations due to the \duckdb-\pd interop overhead. To explore the full potential,  we further modify \duckdb internals to support column swaps between \duckdb tables. We update residuals by swapping the existing residual column with the newly created one, making \sys $1.1\times$ faster than \lgbm.  \revise{
Lastly, we simulate column swap on a commercial DBMS, and see a potential $15\times$ improvement.  
Such modifications, being both feasible ($<100$ LOC on \duckdb) and effective, provide direction for other closed-source columnar DBMSes to support in-DB gradient boosting efficiently}.

\revise{Finally, we apply optimizations to further improve \sys performance.  Algorithmically, we enhance prior batch optimization~\cite{schleich2019layered} by sharing intermediate results (materialized as tables in DBMSes) across batches (tree nodes), leading to a $3\times$ improvement. System-wise,  we leverage inter-query parallelism among SQL queries, accelerating the gradient boosting training by 28\% and random forest by 35\%.}
We conduct extensive experiments with  \sys on various  DBMS backends (local and cloud), using datasets with varying feature numbers, sizes (TPC-DS with SF $10{\rightarrow}1000$), and join graph complexity (galaxy schemas), against SOTA ML libraries (\lgbm,\xgb,\texttt{Sklearn}) and In-DB systems (\lmfao,\mad).
\textbf{On a single node, \sys trains a gradient boosting model with 100 trees $1.1\times$ faster than \lgbm, and is $3\times$ faster for random forests. 
\revise{On multiple nodes, \sys outperforms the Dask~\cite{rocklin2015dask} versions of \lgbm and \xgb by ${>}9\times$.}
\sys easily scales in the \# of features, join graph complexity, and dataset size (TPC-DS SF=1000 in \Cref{sec:scala}), whereas \lgbm encounters memory limitations.}

\vspace{-1mm}
\stitle{Note:} The paper is self-contained and references to appendices can be skipped, or can be found in the technical report~\cite{tech}.

\vspace{-3mm}

\section{Related Work}\label{sec:related}

\stitle{Tree-based ML Libraries.}  Random forest~\cite{breiman2001random} and gradient boosting~\cite{friedman2002stochastic} are the de-facto ensemble models supported by almost all standard ML libraries including \texttt{Sklearn}~\cite{scikit-learn}, \texttt{TensorFlow}~\cite{tensorflow2015-whitepaper} and \texttt{Keras}~\cite{chollet2015keras}. The leading Tree-based ML libraries are \lgbm~\cite{ke2017lightgbm} and \xgb~\cite{chen2016xgboost}, both of which are highly optimized and outperform others as per previous benchmarks~\cite{anghel2018benchmarking}. According to Kaggle 2021 surveys~\cite{kaggle_survey}, among all the commonly used ML libraries, \xgb and \lgbm are ranked $4^{th}$ and $6^{th}$ respectively for popularity (while all the top 3 also support Tree-based ML). However, none of them apply factorized ML for normalized datasets.

\stitle{In-DB ML systems.} 
Most in-DB ML works~\cite{hellerstein2012madlib,feng2012towards,stonebraker2011architecture} focus on extending DBMSes (e.g., PostgreSQL for \mad) to support linear algebra using UDFs and user-defined types.  However, these are not needed for tree-based models that only rely on simple aggregations.  

Other work optimizes ML training by leveraging specialized features (distributed execution~\cite{jankov2019declarative,jankov2021distributed,xin2013shark,boehm2016systemml,kraska2013mlbase} or GPU acceleration~\cite{asada2022share,hu2021tcudb,paul2021database,li2017mlog}) of a specific DBMS.   Although we focus on optimizing the single-node CPU setting via SQL rewrites, \sys can run on any DBMS and benefit from its optimizer and executor.  For instance, \Cref{sec:scala} scales decision tree training on TPC-DS SF=1000 using a cloud DBMS. 
We leave further DBMS-, data-, and model-specific optimizations to future work.  

In practice, cloud vendors (Azure~\cite{azureML}, Redshift~\cite{redshiftML}, BigQuery~\cite{mucchetti2020bigquery}, Snowflake~\cite{snowflakeML}) offer SQL syntax to train tree models.  Under the covers, however, they still materialize and export join results to an external library (\lgbm or \xgb).   Thus, we do not consider them in-DB ML from a performance or portability standpoint.

\stitle{Factorized ML systems.} 
Factorized ML is optimized to train models over normalized databases. 
There are two classes. The first class translates ML models as aggregations over an appropriately designed semi-ring, and pushes aggregations through joins to achieve asymptotically lower time complexity. They support many popular models (ridge regression~\cite{schleich2019layered}, SVM~\cite{khamis2020functional}, factorization machine~\cite{schleich2019layered}), and approximate others (k-means~\cite{curtin2020rk}, GLM~\cite{huggins2017pass}).
Of these, only \lmfao~\cite{schleich2019layered} supports decision trees, albeit in a limited way (see \Cref{exp:lmfao}).  Further, most factorized ML works build specialized query optimizers and executors from scratch, which hinders portability (of the systems and performance wins)  to existing DBMSes.  
\sys extends factorized ML to tree models via vanilla SQL queries, and shows competitive performance with \lgbm.

The second class factorizes linear algebra operations over joins~\cite{chen2016towards,yang2020towards,kumar2015learning} by carefully iterating over the tables, caching intermediates, and combining their results. Compared to the first class, these works support any model that can be formulated as linear algebra statements, such as K-means and GLM (without approximation). However, they only reduce space and not time complexity, and are worse than naive denormalization when data fits in memory~\cite{kumar2015learning}.

\vspace{-2mm}
\section{Background}\label{s:background}

We provide the background of factorized  tree-based models.

\begin{table*}
\begin{center}
\setlength{\tabcolsep}{0.4em} 
\begin{tabular}{c c c c c} 
  \textbf{Semi-ring} & \textbf{Zero/One} & \textbf{Operator} & \textbf{Lift} \\ 
  \hline
  \thd{
Variance \\
$(\mathbb{Z},\mathbb{R},\mathbb{R})$
} & 
\thd{{\bf 0:} $(0,0, 0)$ \\
{\bf 1:} $(1,0, 0)$ }
&
\thd{$(c_1, s_1, q_1) \oplus (c_2, s_2, q_2) = (c_1 + c_2, s_1 + s_2, q_1 + q_2)$\\
$(c_1, s_1, q_1) \otimes  (c_2, s_2, q_2) = (c_1 c_2, s_1 c_2 + s_2 c_1, q_1 c_2 + q_2 c_1 +2 s_1 s_2)$}
 & 
$(1,y,y^2)$  \\ 
\hline
\thd{
Class Count\\
$(\mathbb{Z}, \mathbb{Z}, ..., \mathbb{Z})$
}
&
\thd{
{\bf 0:} $(0, ..., 0)$ \\
{\bf 1:} $(1, ..., 0)$ 
}
& 
\thd{$(c_1, c_1^1, ..., c_1^k) \oplus (c_2, c_2^1, ..., c_2^k) =(c_1 + c_2 , c_1^1 + c_2^1, ..., c_1^k + c_2^k)$\\
$(c_1, c_1^1, ..., c_1^k) \otimes (c_2, c_2^1, ..., c_2^k) =(c_1  c_2 , c_1^1 c_2 + c_1 c_2^1, ..., c_1^k c_2 + c_1 c_2^k)$}
 & 
$(1, 0, ... , c^y = 1, ..., 0)$ \\ 
\hline
\end{tabular}

\end{center}
\caption{Example commutative semi-rings for decision trees. Variance semi-ring supports regression criteria like Reduction in Variance; Class Count semi-ring supports classification criteria like Gini Impurity, Information Gain, and Chi-Square.}
\vspace*{-8mm}
\label{table:semiring}
\end{table*}
\vspace{-3mm}
\subsection{Annotated Relations and Message Passing} \label{s:backgroundmsgpassing}
This section provides an overview of annotated relations and message passing fundamental to factorized query execution~\cite{abo2016faq,olteanu2015size}.

\stitle{Data Model.}  We use the traditional relational data model with the following notations: Given relation $R$, let uppercase $A$ be an attribute, $dom(A)$ be its domain, $S_R=[A_1,\cdots,A_n]$ be its schema, $t\in R$ as a tuple of $R$, and $t[A]$ be tuple $t$'s value of attribute $A$. For clarity, we include the schema in the square bracket followed by the relation $R[A_1,\cdots,A_n]$. The domain of $R$ is the Cartesian product of attribute domains $dom(R) = dom(A_1)\times\cdots\times dom(A_n)$.

\stitle{Annotated Relations.} The annotated relational model~\cite{green2007provenance,joglekar2015aggregations,nikolic2018incremental} maps each $t\in R$ to a commutative semi-ring $(D, \oplus, \otimes, 0, 1)$, where $D$ is a set, $\oplus$ and $\otimes$ are commutative binary operators closed over $D$, and $0/1$ are the zero/unit elements. These annotations form the basis of provenance semi-rings~\cite{green2007provenance}, and are amenable query optimizations based on algebraic manipulation.  Different semi-ring definitions support different aggregation functions, ranging from standard statistical functions to ML models.  For instance, the natural numbers semi-ring $(\mathbb{N},+,\times,0,1)$ allows for integer addition and multiplication, and supports the \cnt aggregate. For an annotated relation $R$, let $R(t)$ denote tuple $t$'s annotation. 

\stitle{Semi-ring Aggregation Query.} Semi-ring aggregation queries can now be re-expressed over annotated relations by translating group-by (general projection) and join operations respectively into  $+$ and $\times$ operations over the semi-ring annotations:
\vspace{-1mm}
\begin{align}
  (\gamma_\mathbf{A} R)(t) = & \sum \{R(t_1) | \HS t_ 1 \in R , t = \pi_{\mathbf{A}} (t_1 )\} \\
(R\Join T)(t) =& \HS R(\pi_{S_R} (t)) \otimes T(\pi_{S_T} (t)) 
\end{align}
\noindent (1) Each group-by result annotation in $\gamma_\mathbf{A} R$ sums all the annotations in its input group. (2) In $R \Join T$, the annotation of each join result is the product of annotations from corresponding tuples in $R$ and $T$.

\stitle{Aggregation Pushdown.}
The key optimization in factorized query execution~\cite{abo2016faq,schleich2016learning} is to distribute aggregations (additions) through joins (multiplications). 
Consider $\gamma_D (R[A,B] \Join S[B,C] \Join T[C,D])$. The naive execution first materializes the join and then computes the aggregation. This costs $O(n^3)$ where $n$ is each relation's cardinality. An alternative evaluation could apply aggregation (addition) to $R$ before joining (multiplication) with $S$, aggregate again before joining with $T$, and then apply a final aggregation.   The largest intermediate result, and thus the join complexity, is now $O(n)$.  
\vspace{-1mm}
$$\gamma_{D} (\gamma_{C} ((\gamma_{B} R[A,B]) \Join S[B,C]) \Join T[C,D])$$
\stitle{Message Passing.} Given a join graph, the above optimization can be viewed as {\it Message Passing}~\cite{pearl1982reverend}.   While {\it Message Passing} supports general SPJA queries~\cite{cjt}, it is sufficient for tree models to restrict ourselves to SPJA queries with zero ($\gamma$) or one ($\gamma_A$) group-by attribute.
Message passing operates over a tree that spans the join graph.  For the root, we can pick any relation (taking cost models into account) that contains the grouping attribute.   Then we direct all the edges of the join graph toward the root to form a tree\footnote{The algorithm applies to acyclic join graphs.  Cyclic join graphs can be pre-joined into acyclic join graphs using standard hypertree decomposition~\cite{abo2016faq,joglekar2016ajar}.}.

Starting from the leaf, we send messages along its path to the root. Each message is computed as:
(1) Join the relation with all incoming messages from children relations, if any.   This blocks until all children have emitted messages.  Then (2) let $\mathbb{A}$ be the attributes in the current relation that are also contained in the remaining relations in the path to the root.  Compute $\gamma_\mathbb{A}$ over the previous join result, and emit the result to the parent relation.  

Consider again $\gamma_D (R[A,B] \Join S[B,C] \Join T[C,D])$ along the join graph: $R-S-T$. If we choose T as the root, then the directed join graph is $R \rightarrow S \rightarrow T$ and the messages are: 
\vspace{-1mm}
$$m_{R\rightarrow S = \gamma_{B} R[A,B]},\hspace{2em}m_{S\rightarrow T} = \gamma_{C} (m_{R\rightarrow S}\Join S[B,C])$$
Once the root receives and joins with all messages, it performs \textbf{absorption}, which simply applies the final group-by: $\gamma_{D} (m_{S\rightarrow T} \Join T[C,D])$. In some cases, the aggregate result is already part of the semi-ring (e.g., \cnt and the natural numbers semi-ring); in other cases, such as tree-based models, the semi-ring decomposes the training metric into their constituent statistics, so we combine them in the final annotation to restore the metric (see next section).

\vspace{-3mm}
\subsection{Tree-based Models}\label{sec:treebackground}

In this section, we describe the traditional tree-based models.  The algorithms are based on CART~\cite{breiman2017classification},  and its extensions (e.g., bagging and boosting) follow standard ML textbooks~\cite{murphy2012machine}.

\stitle{Decision Tree.} Decision tree maps (predicts) target variable $Y$ from a set of features $\mathbf{X}$. Internally, it maintains a tree structure of selection predicates, with each edge containing a predicate $\sigma$ and selecting data based on the conjunction of predicates along the path from the root. Leaf nodes associate with a prediction value $p \in dom(Y)$. The selection predicates by leaves are mutually exclusive and collectively exhaustive in $Dom(\mathbf{X})$. For $t\in Dom(\mathbf{X})$, decision tree predicts by traversing the edges where $t$ meets the edge predicate until a leaf is reached, then outputs its prediction $p$.

Training a Decision Tree over relation $R$ with features $\mathbf{X}$ and target variable $Y$ where $\mathbf{X} \cup {Y} \subseteq S_R$ involves recursively splitting $R$ to minimize a criterion $c(\cdot)$. For regression, the popular criterion is Variance for root mean square error ($rmse$), while Gini Impurity, Entropy, and Chi-Square are common for classification.  Numerical attribute splits use inequality ($\sigma_{A > v}, \bar\sigma_{A \leq v}$), and categorical attribute splits use equality ($\sigma_{A = v}, \bar\sigma_{A \neq v}$) or set-based ($\sigma_{A \in V}, \bar\sigma_{A \notin V}$) predicates. Given the criteria $c(\cdot)$ and split $\sigma$, the reduction of criteria after the split is $c(R) - (c(\sigma(R)) + c(\Bar\sigma(R)))$. Tree growth could be depth-wise or best-first~\cite{shi2007best}. Depth-wise growth splits the tree node with the least depth, and best-first growth splits the tree node greedily with the largest criteria reduction. Finally, each leaf prediction is the average $Y$ for regression or mode $Y$ for classification.

\revise{\Cref{alg:tree} presents the training algorithm. Given relation $R_\Join$ (with $Y$  encoded in annotations), features $\mathbf{X}$, and the maximum number of leaves,  the algorithm iteratively calls \texttt{GetBestSplit} to find the best next split (L3,8,9). \texttt{GetBestSplit} iterates through all features, evaluates its best split (based on  {\it reduction of criteria}), and returns the best split across all features (L11-16). For best-first growth, a priority queue (L2) sorts leaf nodes descendingly based on {\it reduction of criteria}.   In each iteration, the algorithm splits the best leaf node (L7), creating two tree nodes that partition the parent node. It then finds the best split for each new node and pushes them to the queue (L8,9).   Training ends when the number of leaves reaches the maximum (L5). Leaf node predictions are computed, and the tree is returned (L10).  Note that \Cref{alg:tree} is a minimal decision tree for clarity, and practical decision trees have additional parameters like regularization (\Cref{sec:semiring}). During training, evaluating the best split (L14) is the most computationally expensive.}
\vspace{-2mm}
\begin{algorithm}
\SetKwFunction{DecisionTree}{DecisionTree}
\SetKwFunction{EvaluateBestSplit}{GetBestSplit}
\SetKwProg{Fn}{Function}{}{}
\caption{\revise{Decision tree training algorithm. L14 (underlined) is the most computationally expensive.} }
\label{alg:tree}
\Fn{\DecisionTree{$R_\Join, \mathbf{X}, maxLeaves$}}{
    $(numLeaves,pq)  \gets (1,priorityQueue())$\;
    $(c, \sigma, X, R_\Join, root) \gets \EvaluateBestSplit(R_\Join, \mathbf{X}, null, null)$\;
    $pq.push(c, \sigma, X, R_\Join, root)$\;
    
    \While{$numLeaves {+}{+}< maxLeaves$}{
        // get the best split with the highest $c$ among leaves\;
        $(\_,\sigma_{parent}, X, R, parent) \gets pq.pop()$\;
        $pq.push(\EvaluateBestSplit(R, \mathbf{X}, \sigma_{parent}, parent))$\;
        $pq.push(\EvaluateBestSplit(R, \mathbf{X}, \lnot\sigma_{parent}, parent))$\;
    }

    \Return{$addPrediction(R_\Join, root)$}
}
\Fn{\EvaluateBestSplit{$R, \mathbf{X}$, $\sigma$, parent}}{
    $(c^*,\sigma^*,X^*,node) {\gets} 0, null, null,createNode(\sigma,parent)$\;
    \For{each feature $X \in \mathbf{X}$}{
        \small \underline{$(\sigma{,}c){\gets}$ best split and criteria reduction for $X$ over $\sigma(R)$}\;
        \lIf{$c > c^*$}{$(c^*, \sigma^*, X^*) \gets (c, \sigma, X)$}
    }
    \Return{$(c^*, \sigma^*, X^*, \sigma(R), node)$}
}
\end{algorithm}

\vspace{-2mm}
\stitle{Bagging and Boosting.} Big, deep decision trees risk overfitting. Ensemble methods combine multiple smaller or shallower decision trees to create a more robust model. Popular tree-based ensemble models include random forests~\cite{breiman2001random} (bagging) and gradient boosting~\cite{friedman2002stochastic} (boosting). Random forests parallelly train decision trees on samples of $R$ and features and aggregate their predictions (e.g., average) for the final result.  Gradient boosting sequentially trains decision trees based on preceding trees' residuals.

\subsection{Factorized Decision Tree}
\label{sec:factorizedtreeback}
We now introduce factorized decision trees over joins. \revise{The training process follows \Cref{alg:tree}, but we optimize the computation of criteria reduction (L14), the most computationally intensive part, by avoiding join materialization.}
We consider a database with a set of relations $\mathcal{R} = \{R_1, R_2, ..., R_n\}$, and aim to train a decision tree over $R_{\Join} = R_1 \Join R_2 ... \Join R_n$ with  a set of features $\mathbf{X}$ and target variable $Y$.  Let $R_Y$ be the relation that contains $Y$ (or pick one if $Y$ is in many relations as a join key).   For simplicity, we will assume natural join, set semantics, and the Variance split criterion; \Cref{sec:semiring} describes extensions to theta and outer joins, bag semantics, and formulas for classification semi-rings.
Factorized learning avoids materializing $R_{\Join}$ by expressing the criterion as semi-ring aggregation queries, and applying aggregation pushdown over join. We first describe semi-ring annotations for trees and then the computation of criterion.

\begin{figure}
\centering
\begin{subfigure}[t]{0.38\textwidth}
     \centering
     \includegraphics[width=\textwidth]{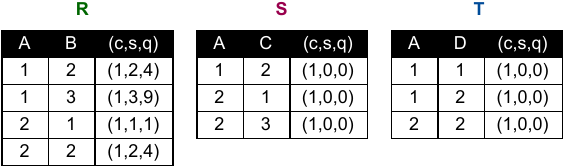}
     \vspace*{-6mm}
      \caption{Relations annotated with variance semi-ring, join graph $R-S-T$, target variable B and features C,D. }
      \label{fig:annotated_relation}
 \end{subfigure}
 
 \begin{subfigure}[t]{0.34\textwidth}
     \centering
     \includegraphics[width=0.9\textwidth]{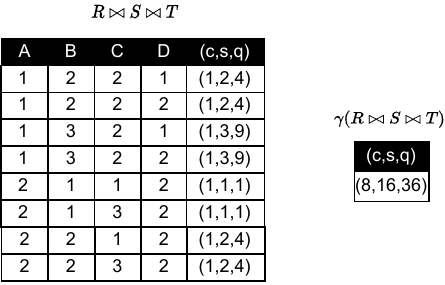}
     \vspace*{-2mm}
      \caption{Naive join $R_\Join$ and aggregation query $\gamma(R_\Join)$.}
      \label{fig:naive_join}
 \end{subfigure}

 \begin{subfigure}[t]{0.33\textwidth}
     \centering
     \includegraphics[width=\textwidth]{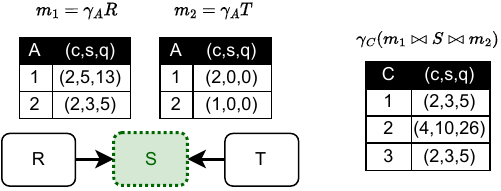}
     \vspace*{-6mm}
      \caption{Message passing for aggregation query $\gamma_C(R_\Join)$. The root node is dotted in green.}
      \label{fig:message_passing}
 \end{subfigure}

 \begin{subfigure}[t]{0.41\textwidth}
     \centering
     \includegraphics[width=0.8\textwidth]{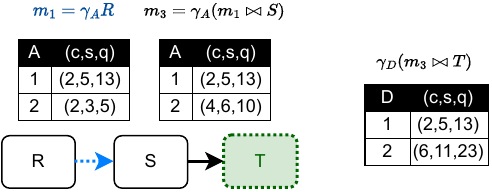}
     \vspace*{-2mm}
      \caption{Message sharing between aggregation queries $\gamma_D(R_\Join)$ and $\gamma_C(R_\Join)$. The reusable message is dotted in blue.}
      \vspace*{-3mm}
      \label{fig:message_sharing}
 \end{subfigure}
 \hfill
 \caption{Factorized decision stump training.}
 \vspace*{-5mm}
 \label{fig:factorizedstump}
\end{figure}

\stitle{Tree Semi-rings.} 
We now illustrate how to use variance semi-ring (\Cref{table:semiring}) to compute the {\it Variance} for regression.  
Each semi-ring defines a $lift(\cdot)$ function~\cite{nikolic2018incremental} that annotates a base tuple with its appropriate semi-ring element.  Variance semi-ring lifts $t\in R_Y$ with $(1,t[Y],t[Y]^2)$, and $t$ from the remaining relations with the $1$ element $(1,0,0)$.    During message passing, the annotations are combined via $\oplus$ and $\otimes$ as defined in \Cref{table:semiring}.  The aggregated semi-ring $\gamma(R_\Join)$ then forms a 3-tuple $(C,S,Q)$ that denotes  the count, sum of $Y$, and the sum of squares of $Y$.   The variance statistic can be derived from this aggregated semi-ring as $variance = Q - S^2/C$. Thus, any filter aggregation query over the join graph can be expressed using message passing and lightweight post-processing.

\begin{example}
\label{exp:variance}
Consider $\mathcal{R} = \{R, S, T\}$ in \Cref{fig:annotated_relation} with variance semi-ring, join graph $R-S-T$ and target variable $Y = B$. To compute the variance over $R_{\Join}$, the naive solution is to materialize $R_{\Join}$ (\Cref{fig:naive_join}), then compute the variance $({=}4)$ over $B$. Instead, we compute $(C,S,Q) = \gamma(R \Join S \Join T) = (8,16,36)$ and $variance {=} Q {-} S^2/C {=} 4$.
\end{example}

\revise{To find the best split, we similarly use variance semi-ring compute the {\it reduction in variance}. Given a tree node with total count ($C_{total}$) and sum of $Y$ ($S_{total}$) over $R$ as precomputed constants, a split $\sigma{:}{f=v}$ for a categorical feature $f$ and a value $v$ yields a variance reduction calculated as $- S_{total}^2/C_{total} + S_\sigma^2 / C_\sigma + (S_{total} - S_\sigma)^2/(C_{total}-C_\sigma)$. Here, $C_\sigma$ and $S_\sigma$ denote the count and sum of $Y$ over $\sigma(R)$, respectively. The optimal split maximizes this variance reduction, evaluated per feature.
For numerical features, we further require prefix sums of $C,S$ (ordered by feature value) to evaluate predicates with inequality (e.g., $\sigma={f<v}$); SQL window functions can achieve this. We provide an example SQL to find the best split, and the full derivation is in \Cref{sec:dt_semiring} due to space limit.}
\begin{example}
\revise{Given relation $R$ and numerical feature $A$. To find its best split point $v$ (i.e., $\sigma{:}{A<v},\lnot\sigma{:}{A\geq v}$), we use the following SQL:}
{\footnotesize
\begin{verbatim}
SELECT A, -({$stotal}/{$ctotal}) * {$stotal} + (s/c) * s
       + ({$stotal} - s)/({$ctotal} - c) * ({$stotal} - s) AS criteria
FROM (SELECT A, SUM(c) OVER(ORDER BY A) as c, SUM(s) OVER(ORDER BY A) as s
    FROM (SELECT A, sum(Y) as s, COUNT(*) as c FROM R GROUP BY A))
ORDER BY criteria DESC LIMIT 1;
\end{verbatim}
}
\end{example}

\vspace{-1mm}
\stitle{Message Caching.} 
Training a decision stump requires computing the set of aggregation queries grouped by each feature: $\{\gamma_X(R_\Join) | X \in \mathbf{X}\}$.    Executing each independently via message passing  is wasteful due to reusable messages.   Consider the example in \Cref{fig:factorizedstump}:
\vspace{-1mm}
\begin{example} \label{exp:groupby} 
Let $\mathbf{X} = \{C,D\}$, and suppose we first aggregate on $C$ (\Cref{fig:message_passing}). We choose S as the message passing root because it contains $C$; we then pass messages $m_1$ from R to S and  $m_2$ from T to S, and absorb the messages into $S$.  We then aggregate on $D$ (\Cref{fig:message_sharing}).  We choose T as the message passing root, pass $m_1$ from R to S,  $m_3$ from S to T, and absorb into T. The two queries can reuse $m_1$.
\end{example}
\vspace{-1mm}
Recent work~\cite{cjt} shows that a simple caching scheme that materializes all messages between relations in the join graph (in both directions) is effective in various  analytical workloads.  The factorized learning system \lmfao~\cite{schleich2019layered} also optimizes batch queries for splitting a single decision tree node.  For decision trees, where each query groups by ${\leq} 1$ feature, its optimizations are equivalent to this simple caching scheme. 
However, \lmfao optimizes a single decision tree node's batch of queries, missing work-sharing across tree nodes and not supporting residual updates for boosting.

\vspace{-2mm}
\section{Factorized Gradient Boosting}
\label{sec:boosting}

This section delves into the algorithms for factorized gradient boosting, which iteratively trains the next tree to predict the residuals from preceding trees. 
Recall from \Cref{sec:treebackground} that, each leaf $l$ in the decision tree is associated with a predicate $l.\sigma$ and prediction $l.p$. We aim to update, for each leaf $l$, the target variable of $t\in l.\sigma(R_\Join)$ to the residual $t[\mathcal{E}]{=}t[Y]{-}l.p$. 
{\it The fundamental challenge is to train decision tree over the updated $Y$ in $R_{\Join}$ without materializing $R_{\Join}$.}

One tempting approach is to treat this as a view update problem: we  update $Y$ in the  $R_\Join$ view, translate it into updates over base relations,  re-lift the updated base relations, and train the next factorized decision tree.
However, view updates are susceptible to  "side-effects"~\cite{kotidis2006updates}: for $t\in R_{Y}$,  $y=t[Y]$ can be duplicated in $R_\Join$ due to 1-N (or M-N) joins.
If one duplicate is updated to a new value $y'$, updating $t[Y] = y'$ in $R_{Y}$ would inadvertently cause other duplicates to be updated as well (as "side-effects").
To address it, new tuples are created in base relations for new mappings.
These "side-effects" are particularly problematic for "residual updates", as the full $Y$ column in $R_\Join$ is updated, requiring $O(|R_\Join|)$ new tuples.

\revise{
We address these issues for snowflake (a single fact table) and galaxy schemas (multiple fact tables)---arguably the most common database schema forms.
Snowflake schemas exhibit a 1-to-1 relationship between the fact table $F$ and $R_\Join$, and we exploit this to efficiently updates $Y$ over $F$.  We then discuss the system challenges when performing full-column updates.
These techniques do not apply to galaxy schemas due to the M-N relationships between the fact tables.  While directly updating $Y$ in $R_\Join$ remains challenging, we identify the {\it addition-over-multiplication preserving} property that allows us to directly update semi-ring aggregates derived from updated $Y$, and use this to efficiently support the popular $rmse$ criteria}.
\vspace{-3mm}
\subsection{Snowflake Schemas}
\label{sec:snowflake}
Snowflake schemas have one fact table $F$ and N-to-1 paths along the dimension tables; this means that $F$ is 1-1 with $R_{\Join}$\footnote{We assume no missing join keys, or use left outer join to maintain 1-to-1 relationship.} and we can directly update $F$. 
Let us first assume that $Y$ is in $F$ (i.e., $R_Y=F$).
The main challenge is to translate a leaf $l$'s predicate $l.\sigma$, which may refer to dimension attributes, into one over $F$.  We do so by translating $l.\sigma$ as semi-join predicates\footnote{We treat left semi-joins as filters over the left relation so  its annotations don't change. } over $F$. Given join path $F-D_1-...-D_k$, and $\sigma(D_i)$, we ``move'' the predicate to be $\sigma'(D_{i-1})$:
$$D_{i-1} \Join \sigma(D_i)  = (D_{i-1}\ltimes \sigma(D_i)) \Join D_i = \sigma'(D_{i-1}) \Join D_i$$
\revise{where $\sigma'$ is over the join keys $\mathcal{J} = S_{D_{i-1}} \cap S_{D_{i}}$. Note that $\sigma$ doesn’t have
to be equality-based but can be of any arbitrary type: $\sigma$ is applied to $\pi_\mathcal{J}(\sigma(D_i))$ to identify the matching join keys for the semi-join. }
If $Y$ is in a dimension table $D$ (i.e., $R_Y\neq F$), we join the relations along the path from $F$ to $D$ and project all attributes in $F$ along with $Y$.   This reduces to the first case above.

\stitle{System Challenge:} 
\revise{Since $|F|=|R_\Join|$,  factorized gradient boosting {\it does not offer an asymptotic advantage over specialized ML libraries}.  Theoretically, if the performance is comparable, it could still benefit from the parallelization, scaling, and administrative features of DBMSes.  However, in practice, we find that bulk updates to $F.Y$ are a major bottleneck (\Cref{sec:pilot}) for existing DBMSes.   The next section dives into these system challenges, and explores system design, logical, and physical optimizations to address them.}

\vspace{-2mm}
\subsection{Galaxy Schemas}
\label{sec:galaxy}
\revise{Galaxy schemas model M-N relationships between multiple fact tables. 
They are prevalent in enterprise settings~\cite{saxena2014data}; the recent ``semantic layers'' trend~\cite{chatziantoniou2020data,dbtdatamodel,lookerdatamodel,weighing} (pre-defined denormalized views) have made analysis and ML over galaxy schemas more accessible.   However, galaxy schemas induce the 1-N relationship between $R_Y$ and $R_{\Join}$ that causes side-effects~\cite{kotidis2006updates} during residual updates.}

\revise{We observe that individual $Y$ values are not needed for training---the split criteria only refers to {\it semi-ring aggregates} over the updated $Y$ values (\Cref{sec:factorizedtreeback}), and can potentially be updated efficiently.
For instance, for a leaf $l$ with original aggregates $c=\sum 1, s=\sum y$ over $l.\sigma(R_\Join)$,
we can directly update the sum of residuals as $\sum (y - l.p) = \sum y - \sum l.p = s- c \times l.p$, without referencing individual $Y$ values.}

\revise{Unfortunately, this approach is not efficient for arbitrary split criteria.
To review, factorized ML maps (via $lift(\cdot)$) $y$ in the real number semi-ring $\mathbb{R}(\mathcal{R}, +, \times, 0, 1)$ (for base relation) to another semi-ring $\mathbb{S}(\mathcal{S}, \oplus, \otimes, 0', 1')$ (e.g., variance semi-ring for $rmse$), and computes $\sum lift(y)$ over $R_\Join$ with aggregation pushdown.
To directly update aggregates given leaf $l$, we want to find a function $f{:} (\mathcal{S}, \mathcal{R}) \rightarrow \mathcal{S}$, such that $f$ takes the original aggregates $\sum lift(y)$ (without referencing individual $y$) and additive inverse of prediction $l.p$\footnote{Additive inverse of $l.p$ exists because $\mathbb{R}$ is also a ring.} as input, and outputs updated aggregates: $f(\sum lift(y),-l.p) = \sum lift(y-l.p)$.
To maintain our asymptotic benefits, both $f$ and semi-ring operation/annotation should be of constant time/size. }

\revise{Such $f$ does not exist for all semi-rings. For example, {\it mean absolute error} ($mae$) relies  on the count and sum of signs: $\sum 1, \sum sign(y)$. The naive semi-ring would track $\sum lift(y) = (\sum 1,\sum sign(y))$. However, $f$ doesn't exist because given the leaf $l$, $\sum sign(y-l.p)$ cannot be solely decided by $\sum 1,\sum sign(y), -l.p$, as it depends on individual $y-l.p$. A naive solution uses a semi-ring that collects all $y$ values, but its annotation size will be $O(|R_\Join|)$ and defeats the purpose of factorization. 
To this end, we identify a sufficient property for constant sized semi-ring $\mathbb{S}$ to construct $f$:}
\vspace{-2mm}
\begin{definition}{Addition-to-Multiplication Preserving.}
\label{update_property}
\revise{Given two semi-rings $\mathbb{R}(\mathcal{R}, +, \times, 0, 1)$, $\mathbb{S}(\mathcal{S}, \oplus, \otimes, 0', 1')$, a lift function $lift(\cdot): \mathcal{R} {\to} \mathcal{S}$ is addition-to-multiplication preserving from $\mathbb{R}$ to $\mathbb{S}$ if \\$lift(d_1 + d_2) = lift(d_1) \otimes lift(d_2)$ for any $d_1, d_2 \in D$.}
\end{definition}
\vspace{-2mm}
\revise{Based on the property, we can construct $f$ as:}
\vspace{-2mm}
\begin{align*}
f\left(\textstyle\sum lift(y),-l.p\right) & = \left(\textstyle\sum lift(y)\right) \otimes lift(-l.p) \quad (f\text{{ definition}} ) \\
& = \textstyle\sum \left(lift(y)\otimes lift(-l.p)\right) \quad (\otimes\text{{ distributive}} ) \\
& = \textstyle\sum lift(y-l.p) \quad \text{{(Add-to-Mul Preserving)}}
\end{align*}
\revise{It is easy to verify that, the variance semi-ring (and lift) in \Cref{table:semiring} satisfies the property:  $lift(y-l.p) = (1, y -l.p, l.p^2 + y^2 -2l.p \times y) = (1, y, y^2) \otimes (1, -l.p, l.p^2)=  lift(y) \otimes lift(-l.p)$. 
In contrast, there is no known $lift$ and constant-sized semi-ring with this property for $mae$.  \Cref{sec:limitation} describes a possible future direction using polynomial approximations~\cite{huggins2017pass} of the semi-ring aggregates. To keep the text concrete, the subsequent text will refer to the $rmse$ criteria}.

\vspace{-2mm}
\subsubsection{Update Relation:} 
\begin{figure}
\centering
\begin{subfigure}[b]{0.13\textwidth}
     \centering
     \includegraphics[width=0.8\textwidth]{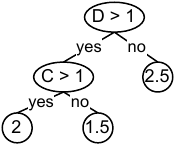}
     \vspace*{-3mm}
      \caption{Decision Tree.}
      \label{fig:decisiontree}
 \end{subfigure}
 \hfill
  \begin{subfigure}[b]{0.16\textwidth}
     \centering
     \includegraphics[width=\textwidth]{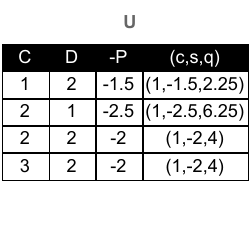}
     \vspace*{-12mm}
      \caption{Update Relation U.}
      \label{fig:update}
 \end{subfigure}
 \hfill
 \begin{subfigure}[b]{0.13\textwidth}
     \centering
     \includegraphics[width=0.8\textwidth]{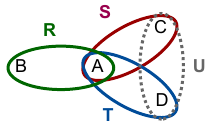}
     \vspace*{-3mm}
      \caption{Join Graph.}
      \label{fig:joingraph}
 \end{subfigure}

     \begin{subfigure}[t]{0.45\textwidth}
     \centering
     \includegraphics[width=0.5\textwidth]{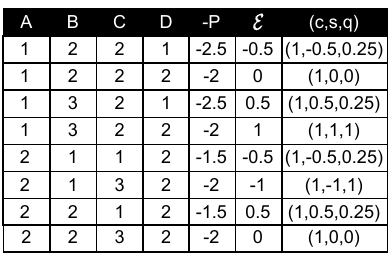}
     \vspace*{-1mm}
      \caption{Materialized $R_\Join$ with prediction additive inverse  $-P$, residual $\mathcal{E} = Y - P$ and annotation $(1, \epsilon,\epsilon^2)$ lifted on $\mathcal{E}$ for the next tree.}
      \vspace*{-4mm}
      \label{fig:residual}
 \end{subfigure}
 \hfill

\caption{Update Relation U for decision tree residual updates over the non-materialized $R_\Join$. Each tuple in $U \Join R_\Join$ has the same annotation as the materialized $R_\Join$ lifted on $\mathcal{E}$. Therefore, the materialization of $R_\Join$ can be avoided.}
\vspace*{-5mm}
\end{figure}

\revise{For decision tree and each of its leaf $l$, we  apply the corresponding $f$ that multiplies aggregates over $l.\sigma(R_\Join)$ with $lift(-l.p)$.
We model this as an {\it Update Relation} $U$  joins with $R_\Join$.
$U$ is constructed as follows: let $\mathbf{A}$ be the set of attributes referenced by $l.\sigma$ for any leaf $l$, and $U$ be the projection $\pi_\mathbf{A}(R_\Join)$, along with a column $-P$ of the additive inverse of leaf prediction (unique as leaf predicates are non-overlapping).
Naively, the next boosting lifts the residual using $lift_{Y-P}(R_\Join{\Join}U)$\footnote{When  $lift$  is applied to a column $C$, we denote it as $lift_C$.}, where $R_\Join{\Join}U$ has to be materialized. Instead, we rewrite this into $lift_{Y}(R_\Join){\Join}lift_{-P}(U)$}:
\vspace{-2mm}
\begin{proposition}
\label{pro:variance}
\revise{Let $t {\in} U {\Join} R_\Join$. $lift_\mathcal{E}(U {\Join} R_\Join)(t) = (lift_Y(R_\Join) \\ \Join lift_{-P}(U))(t)$ if  $lift(\cdot)$ is addition-to-multiplication preserving}.
\end{proposition}
\vspace{-4mm}
\begin{proof}
\revise{Consider any $t {\in} U {\Join} R_\Join$ with $y = t[Y]$ and $-p=t[-P]$. Naively, we materialize $\mathcal{E}=Y-P$, then lift $lift_\mathcal{E}(U \Join R_\Join)(t) = lift(y-p)$.
Instead, we compute $(lift_Y(R_\Join) \Join lift_{-P}(U))(t) = lift(y)\otimes lift(-p)$ without materializing $\mathcal{E}$. They are equivalent because $lift(\cdot)$ is addition-to-multiplication preserving}.
\end{proof}
\vspace{-2mm}
\begin{example}
The decision tree in \Cref{fig:decisiontree} has three leaf nodes: $(\sigma_{D\leq 1}, p{=}2.5)$, $(\sigma_{D>1 \wedge C\leq 1}, p{=}1.5)$,  $(\sigma_{D > 1 \wedge C>1}, p{=}2)$. The set of referenced attributes is $\mathbf{A} {=} \{C,D\}$. The update relation $U$ is shown in \Cref{fig:update}. The semi-ring annotations lifted on $\mathcal{E} {=} Y {-} P$ over the materialized join are shown in \Cref{fig:residual}. It is easy to verify that they  are the same as those in $U \Join R_\Join$, without materialization $R_\Join$.
\end{example}
\vspace{-2mm}

\vspace{-2mm}
\subsubsection{Algorithmic Challenge.}
\label{sec:predtree}
Unfortunately, $U$ can introduce cycles in the join graph over time  (e.g., \Cref{fig:joingraph} shows cycle $A\to C\to D$); whereas {\it Message Passing} requires acyclic join graphs.  Standard hypertree decomposition~\cite{abo2016faq,joglekar2016ajar} removes cycles by joining the relations in a cycle, materializing their join result $R'$ (e.g., $S[A,C]\Join T[A,D]\Join U[C,D]$), and replacing these relations in the join graph with $R'$. However, as the number of trees in model increases, the number of referenced attributes is likely to span the entire join graph---$|U|$ will thus converge to $|R_\Join|$.

To address this, we propose {\it Clustered Predicate Tree} (CPT): each boosted tree restricts split on features that can be pushed to the same fact table. To do so, we cluster relations such that within each cluster, a single fact table $F$ maintains N-to-1 relationships with all other relations. Predicates in this cluster can be rewritten as semi-joins to the same fact table $F$ (\Cref{sec:snowflake}) and won't create cycles~\cite{abo2016faq}.  During training, while the root decision tree node can split on any feature, subsequent splits are confined to attributes within the same cluster. 
Although this might affect model accuracy, it allows efficient residual updates over join graphs, which would otherwise be untrainable using existing techniques.

\begin{figure}
\centering
     \centering
     \includegraphics[width=0.3\textwidth]{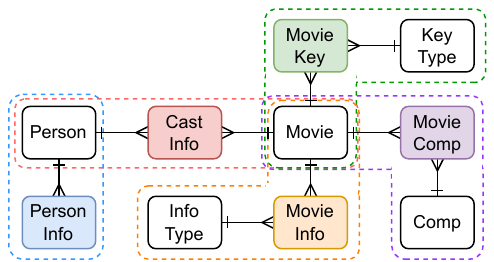}
     \vspace*{-3mm}
\caption{Clusters for IMDB dataset. Each cluster is enclosed by dotted lines and its fact table is filled. }
\vspace*{-6mm}
\label{fig:cluster}
\end{figure}

\vspace{-2mm}
\begin{example}
\Cref{fig:cluster} shows the  join graph of IMDB datasets~\cite{imdb}, which was previously prohibitive to train gradient boosting due to the large join size ($R_\Join$ is ${>}1TB$).  The five clusters are enclosed by  dotted lines and the cluster's fact table is highlighted. If the current tree initially splits on Person's age (in Person Info),  the rest of the tree can only split on attributes in Person or Person Info. 
\end{example}

\vspace{-2mm}
\section{\sys Overview and Optimizations}\label{sec:system}

We now present the system overview and optimizations.
\vspace{-2mm}
\subsection{\sys Developer Interface}

As described in \Cref{s:intro}, our design goals are portability, performance, and scalability.   We evaluate performance and scalability in the experiments, so focus on portability below.

\begin{figure}
\centering
     \centering
     \includegraphics[width=0.4\textwidth]{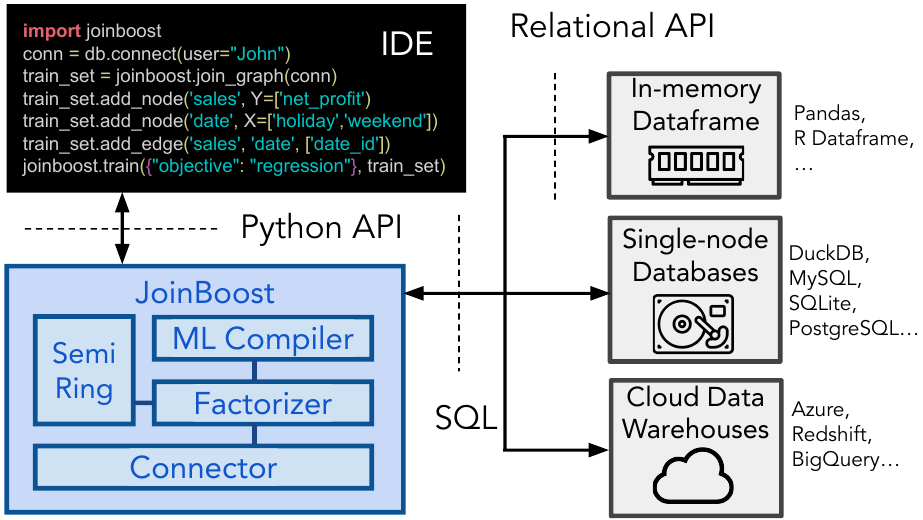}
     \vspace*{-4mm}
\caption{\sys Architecture. \sys translates its Python API calls into message passing algorithms that are executed as SQL queries on backend DBMSes and dataframes. }
\vspace*{-5mm}
\label{fig:arch}
\end{figure}

\stitle{Portability:}  Our design deviates from prior factorized ML works~\cite{schleich2019layered,ke2017lightgbm,chen2016xgboost,khamis2018ac}, which build (fast) custom execution engines.  In contrast, \sys is implemented as a Python library that transparently generates SQL queries to the user's DBMS backend (\Cref{fig:arch}).

\stitle{API Compatibility:} For usability, \sys offers Python API similar to \lgbm and \xgb, and returns models identical to \lgbm. In these libraries, users first define a training dataset and pass it, along with training parameters (e.g., objective, criteria, number of leaves), to a \texttt{train()} method. Likewise, \sys users define the training dataset as a join graph (the relations and join conditions between them), providing a database connection, and specifying the features and target variable. If only relations are specified, \sys infers the join graph that covers those relations from the database schema and raises an error if the graph is ambiguous (e.g., multiple foreign key references between relations) or requires cross-products. Finally, the user passes the training dataset and training parameters to \texttt{train()}. For consistency, \sys accepts the same training parameters as \lgbm.
\vspace{-2mm}
\begin{example}
    \Cref{fig:arch} illustrates a simple example inspired by TPC-DS.
    The user creates a database connection, and initializes the training dataset as a join graph with relations $[\texttt{sales}, \texttt{date}]$, join attribute \texttt{date\_id}, features $\mathbf{X} = [\texttt{holiday}, \texttt{weekend}]$ and target variable $Y = \texttt{net\_profit}$.    Finally, the user chooses model parameters ($\{objective=regression\}$) and runs \texttt{train} over the training dataset. 
\end{example}
\vspace{-2mm}
\sys internally translates the ML algorithms into \texttt{CREATE TABLE} and \texttt{SELECT} SQL queries. Compared to in-DB systems like \mad~\cite{hellerstein2012madlib}, \sys generates pure SQL and does not require user-defined types or functions.  This enables portability (criteria C2): \sys runs on embedded databases, single-node databases, cloud data warehouses, and even \pd and R dataframes~\cite{relationalapi}.

The compiler fully supports decision trees, random forests, and gradient boosting with all learning parameters that \lgbm supports. Currently, regression with $RMSE$ objective supports galaxy schema with Clustered Predicate Trees; other objectives (e.g., $mae$, $huber$, $softmax...$) require snowflake schema.
We are actively extending capabilities to support pruning, dropout, and early stopping, which build on the techniques in the preceding sections.

\revise{One usability challenge is that \sys end users are typically domain scientists who lack expertise in data access and schema, commonly handled by data engineers.  While addressing this is beyond the scope, we highlight the recent trend of "semantic layer"~\cite{chatziantoniou2020data,dbtdatamodel,lookerdatamodel,tableaudatamodel} as a means to bridge the knowledge gap.
In the "semantic layer," data engineers employ SQL-based recipes to transform and preprocess raw data into tables containing meaningful attributes to domain scientists, and can be queried by  \sys using SQL. However, current "semantic layer" still materializes costly joins, given most downstream applications anticipate a single table.
Integrating \sys with semantic layers to avoid join materialization is a potential avenue for future work.}

\stitle{Safety:}
Training shall never modify user data. To achieve that, \sys creates temporary tables in a specified namespace or with a unique prefix. By default, \sys deletes these tables after training, but users can keep them for provenance or debugging.

\subsection{Architecture Overview}
\label{sec:archOverview}
\revise{The {\it ML Compiler} runs the ML logics (\Cref{alg:tree}) to train tree-based models in Python and translates the most computational intensive L14, which identifies the best split across features for tree nodes, into SQL queries to be executed by DBMS.} At this stage, the compiler treats the join graph as a single "wide" table, with the SQL query operating on this table, and factorization is applied in a later step. After training, it returns a reference to the trained model.

The {\it Semi-ring Library} stores semi-ring definitions and translates math expressions in the compiler-generated queries ($\times$, $+$, $lift$) into SQL aggregation functions. 
$lift(R)$ creates a copy of a base relation $R$ that contains an additional attribute for each component in the semi-ring (e.g., c, s, q in the variance semi-ring).  This also ensures that any update in-place will not modify user data.  
In addition to the variance (for regression)  and the class count (for classification) semi-rings,  \sys implements semi-rings for a wide range of popular objectives; it supports $rmse$ for snowflake and galaxy schemas, and {\it mae, huber loss, fair loss, log loss, softmax} and more for snowflake schemas (see \Cref{sec:semiring} for full list).

The {\it Factorizer} decomposes each aggregation query into message passing and absorption queries.  It also materializes each message as a database table, and re-uses them when possible.  After choosing a node split, the factorizer keeps messages that can be reused in descendent nodes (\Cref{alg:factorizedtree}) and drops the rest.

Finally, the {\it Connector} takes our internal SQL representation, and translates them into the appropriate SQL string or dataframe API calls.  Although DBMSes are notorious for incompatible SQL variants, \sys only uses a subset of SQL that is generally consistent across vendors.  For instance, it generates standard non-nested SPJA queries with simple algebra expressions. 

\vspace{-2mm}
\subsection{Residual Updates Logical Optimization}
Although correct, the residual update technique in \Cref{sec:boosting} is still expensive to implement naively.   For simplicity, if we assume a single join attribute $A$ between the fact table $F$ and update table $U$, and the variance semiring, the SQL query would be:

{\footnotesize
\begin{verbatim}
CREATE TABLE F_updated AS
    SELECT F.c*U.c AS c, F.s*U.c+U.s*F.c AS s, 
           F.q*U.c+U.q*F.c+2*F.s*U.s AS q, ...
      FROM F JOIN U ON A 
\end{verbatim}
}
\vspace{-2mm}
\noindent where $c$, $s$, $q$ are semiring components, and the remaining columns in $F$ are copied over (shown as $\dots$).  Unfortunately, this is ${>}50\times$ slower than \lgbm's residual update procedure (see experiments below), because $U$ can potentially be as large as the materialized $R_\Join$.  To this end, we present an optimization for snowflake and galaxy schemas that completely avoids materializing $U$ as well as $F\Join U$.

\subsubsection{Semi-join Update Optimization.}
\label{sec:semijoinupdate}
We  directly \texttt{UPDATE}  $F$'s semi-ring annotations.  Let us start with a snowflake schema. 
Each decision tree leaf $l$ logically corresponds to a separate join graph containing a set of messages (\Cref{fig:msgsharenodes}), with $l.\sigma$ as its predicate and $l.p$ as its prediction.
Using the semi-join optimization (\Cref{sec:boosting}), we translate predicates over $R_\Join$ (e.g., $l.\sigma$) into semi-joins between $F$ and relevant incoming messages $\mathcal{M}$. A message $m_i\in\mathcal{M}$ is relevant if it is along a join path from a relation containing an attribute in $l.\sigma$ to $F$. For each leaf node $l$, we execute the following query where $F.a_i$ is the join attribute with its relevant incoming message $m_i$:

{\footnotesize
\begin{verbatim}
    UPDATE F SET c * 1 AS c, s - l.p*c AS s, q + l.p*l.p*c - 2*s*l.p AS q
           WHERE F.ai IN (SELECT ai IN mi) AND ...
\end{verbatim}
}
\vspace{-2mm}
In some databases, updates in place can be very slow.  Thus an alternative is to create a new fact table with the updated semi-ring annotations.  Let $l_j$ be the $j^{th}$ leaf in the decision tree, and $m_{i,j}$, $a_{i,j}$ be the $i^{th}$ message and its join attribute with $F$ in $l_j$'s join graph:  

{\footnotesize
\begin{verbatim}
  CREATE TABLE F_updated AS SELECT 
        CASE WHEN F.a_ij IN (SELECT a_ij FROM m_ij) AND ... THEN s - l_j.p*c
          WHEN ... // Other leaves
      END AS s, ...  // other semi-ring components 
        ...  // copy other attributes in F
    FROM F
\end{verbatim}
}
\vspace{-2mm}
\noindent The same ideas apply to galaxy schemas, where $F$ corresponds to the fact table of the current tree's cluster.     Further, we show in the technical report~\cite{tech} that $c$ and $q$ are not necessary to materialize; thus only $s$ is needed for the variance semi-ring.

\subsubsection{Pilot Study.}
\label{sec:pilot}
When should we perform in-place updates as compared to creating new tables?  On which DBMSes?  We now report a microbenchmark to understand the performance trade-offs, and use them to motivate a new optimization.  
We used an Azure VM, with 16 cores, 128 GB memory, and an SSD.

\stitle{Workloads.} We create a synthetic fact table $F(s,d,c_1,...,c_k)$ with $100M$ rows to simulate residual updates in a decision tree with 8 leaves. $s$ is the semi-ring column to update, $d\in[1,10K]$ is the join key, and $c_k$ are simply extra columns that would need to be duplicated in a new table.
For the $i^{th}$ leaf node, its prediction is a random float, and we construct its semi-join message $m_i(d)$ to contain all values in $(1250\times(i-1), 1250\times i]$.

\stitle{Methods.} We evaluate three approaches.
\texttt{Naive} materializes Update Relation $U$, then re-create fact table: $F' = F \Join_\mathbf{A} U$ as discussed in \Cref{sec:boosting}.  
\texttt{SET} and \texttt{CREATE} use the update-in-place and create table optimizations in the preceding subsubsection.  \texttt{CREATE-k} denotes the number of extra columns in $F$, where $k\in\{0,5,10\}$; we set $k=0$ for \texttt{Naive}, and $k$ does not affect \texttt{SET.}

\stitle{DBMSes.} We evaluate two systems. \dbmsx is a popular commercial RDBMS that supports both column-oriented (\texttt{X-col}) and row-oriented (\texttt{X-row}) storage and query processing. \dbmsx is disk-based only, and we set the isolation and recovery to the lowest level (read uncommitted and minimum logging). \duckdb~\cite{raasveldt2019duckdb} is a popular embedded column-oriented OLAP DBMS and is highly performant~\cite{ClickBench}. \duckdb has disk-based (\texttt{D-disk}) and memory-based (\texttt{D-mem}) modes. 
As a reference, we also use \lgbm to train 1 iteration of gradient Boosting with the same training settings, and report residual update time.

\begin{figure}
\centering
\includegraphics[width=0.46\textwidth]{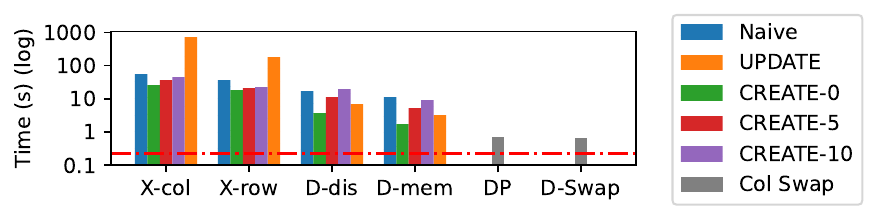}
\vspace*{-4mm}
\caption{Residual update time (log) for different DBMSes using different methods. The red horizontal line is the residual update time for \lgbm. \texttt{Column swap} (\texttt{DP}, \texttt{D-Swap}) achieves competitive residual update performance. }
\vspace*{-5mm}
\label{fig:micro_update}
\end{figure}

\stitle{Experiment Results.} 
\Cref{fig:micro_update} shows that \texttt{Naive} incurs high materialization and join costs.  \texttt{CREATE} is $\sim2\times$ faster for \dbmsx and $\sim4\times$ faster for \duckdb, but its cost grows linearly with $k$. \texttt{SET} mainly depends on the DBMS---it is prohibitive for \dbmsx, but more efficient than \texttt{CREATE} when $k>5$ for \duckdb.  All DBMS approaches take ${>}3s$ for updating residuals. In contrast, \lgbm stores the target variable in a C++ array and performs parallel writes; its residual update takes ${\sim}0.2s$.
These poor results are due to four main factors.   

\begin{myitemize}
\item {\bf Compression:} \texttt{CREATE} in \texttt{X-col} incurs high compression costs: the database is $1GB$ in \texttt{X-col} as compared to $2.6GB$ for \duckdb.   This also penalizes \texttt{SET} due to decompression.  

\item {\bf Write-ahead Log (WAL)} introduces costly disk writes.

\item {\bf Concurrency Control (CC):} In-memory \duckdb doesn't use WAL, but incurs  MVCC~\cite{neumann2015fast} overheads, including versioning, and logging for undo and validation. 

\item {\bf Implementation:} \duckdb's update is currently single-threaded.  
\end{myitemize}

\subsection{Residual Updates Physical Optimization}
\label{sec:duckpandas}

Logical rewrites are effective but still much slower than \lgbm's residual updates, even when existing CC mechanisms are lowered.   We observe that \sys does not need durability and concurrency control, since it writes to private tables, performs application-level concurrency control (\Cref{sec:opt}), and can simply re-run upon failure.   Compression is also unnecessary for the heavily updated columns.  \revise{Unfortunately, WAL, CC, and compression are deeply integrated into DBMS designs, and nontrivial to fully disable\footnote{   \revise{Verified with \duckdb and \dbmsx developers.}}.}

\revise{We note that columnar engines are well suited to avoid these costs by adding the new residual column as a projection~\cite{stonebraker2018c}, but is
not generally supported (e.g., by \duckdb, \dbmsx).  We thus evaluate its benefits using three methods: one that uses \duckdb's existing APIs to swap columns in \pd dataframe, one that adds ``column swapping'' to \duckdb, and one that simulates it in \dbmsx.} 

The first solution,  called \texttt{DP} (\duckdb + \pd), uses the existing \duckdb Relational API~\cite{relationalapi} to directly access \pd dataframes~\cite{reback2020pandas} -- an in-memory data structure that stores columns in contiguous uncompressed C arrays.  \revise{Internally, it uses a custom scan operator for \pd's matrix format, and DuckDB's executor for the rest of the query.}  We store the fact table $F$ as a dataframe, and the remaining tables in \duckdb, and join them via the relational API during training.
\revise{For residual update,  we submit a query over \pd and native relations to compute the semi-ring annotations for the updated residuals, and store the result in a NumPy array.  Then using \pd, we replace the old column in $F$ with the new NumPy array (a pointer swap). This is fast because it avoids WAL or CC overhead, and}
reduces residual updates to $0.72s$--competitive with \lgbm (\Cref{fig:micro_update}).
The drawback is that the scan overhead slows down the join-aggregation query, and increases training time by ${\sim}1.6\times$ as compared to querying only native relations (\Cref{sec:db_exp}).

\revise{The second approach, \texttt{D-Swap}, modifies \duckdb internals slightly (${<}100$ LOC) to support pointer-based column swapping between two \duckdb tables.} \duckdb stores tables in row groups, with each group containing pointers to the column data.
\texttt{D-Swap} iterates through the row groups of the two relations and swaps the column pointers. \revise{Such column swap is a schema-level modification, so it is very fast and side-steps decompression, CC, and WAL overheads. It achieves similar residual update performance to \texttt{DP} (\Cref{fig:micro_update}) without degrading aggregations (\Cref{sec:db_exp}).  }

\revise{Finally, to extend beyond \duckdb, we also simulate column swapping's impact on \dbmsx in \Cref{sec:db_exp}, and see a potential $15\times$ improvement.  We believe that implementing such column swap in closed-source DBMSes is both feasible and effective for residual updates.    We encourage columnar DBMS developers to incorporate this operation for efficient In-DB gradient boosting. }

\subsection{Optimizations and Features}
\label{sec:opt}
We implement various optimizations and features in \sys for performance and usability. Due to space limit, we only discuss the most critical ones here, while the rest can be found in \Cref{app:opt}.

\begin{figure}
\centering
\includegraphics[width=0.4\textwidth]{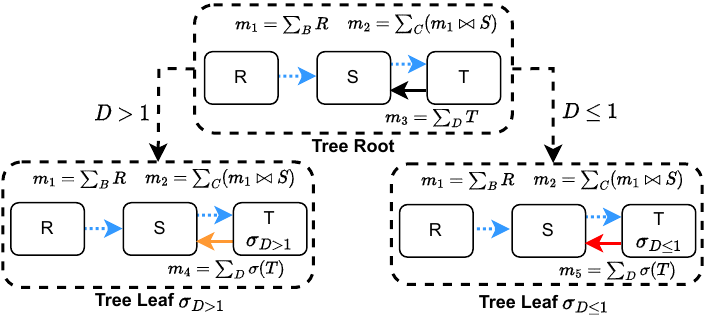}
\vspace*{-5mm}
 \caption{Share computations between the parent node and children nodes after the split. Dotted messages are shared.}
 \vspace*{-6mm}
 \label{fig:msgsharenodes}
\end{figure}

\subsubsection{Message Sharing Among nodes}
\label{alg:factorizedtree}
Previous factorized ML algorithms~\cite{schleich2019layered,khamis2018ac} rely on batch optimization of aggregation queries, suitable for models with closed-form solutions (e.g., ridge regression~\cite{schleich2019layered}). However, tree-based model training is iterative: queries for child nodes depend on the parent node's split, making batching ahead impossible. Consequently, previous algorithms either miss work-sharing opportunities between tree nodes or batch all possible splits, which is impractical due to exponential growth in tree depth.
Our key observation is that messages as intermediate results can also be shared among tree nodes.  Consider the example:

\begin{example}
Following \Cref{exp:groupby}, suppose the best root split is $\sigma_{D > 1}$. The split applies $\sigma$ and  $\lnot\sigma$ to $T$ (as it contains $D$), creating two leaf nodes with relations: $\sigma(R_{\Join})$ and $\lnot\sigma(R_{\Join})$.  Then, for each leaf node, we compute the batch of aggregation queries and identify the next split through {\it Message Passing}. As shown in \Cref{fig:msgsharenodes}, messages (dotted blue) along the path $R\rightarrow S\rightarrow T$ are the same for both leaf and root nodes.  We only need to recompute message $m_{T \rightarrow S}$ in each tree leaf, skipping $m_{S\rightarrow R}$ as $R$'s attributes aren't used in this model.
\end{example}

In general, after a split on $R_i$, all messages along the path to $R_i$ can be re-used. 
This is orthogonal to prior batch optimization work~\cite{schleich2019layered} as we can cache and reuse messages after batching for future nodes, further improving batch optimization by ${>}3\times$ (\Cref{exp:lmfao}).

\subsubsection{Sampling for Random Forest}
A random forest model simply trains multiple decision trees over random samples of the training data and features, and aggregates (e.g., averages) their predictions during inference.  
\revise{Feature sampling can be easily implemented by using a random subset of $\textbf{X}' {\subseteq }\textbf{X}$ features for \Cref{alg:tree}. The main challenge is efficiently sampling over non-materialized $R_{\Join}$: Naively sampling each relation is (1) not uniform and independent, and (2) may produce non-joinable and hence empty samples.}

\revise{To address these, we use previous ancestral sampling~\cite{ch1993sampling,zhao2018random,shanghooshabad2021pgmjoins}:
Given $R_{\Join} {=} R_1 {\Join} R_2 {...} {\Join} R_n$ with total \cnt $C$ (as a constant), ancestral sampling treats  $R_{\Join}$ as the full probability table, with each tuple's probability $1/C$. Then, the marginal probability over a set of attributes $\textbf{A} \subseteq S_{R_\Join}$ is $P(\textbf{A}) = \gamma_{\textbf{A}, count(*)/C}(R_{\Join})$.
Ancestral sampling starts from a root relation (e.g., $R_1$) in the join graph and draws weighted sample $t$ based on its marginal probability (e.g., $P(S_{R_1})$); $P(S_{R_1})$ requires a \cnt semi-ring aggregation query that can be computed without materializing $R_{\Join}$  (\Cref{s:backgroundmsgpassing}). It then traverses the join graph to the next relation (e.g., $R_2$ with the join key $\mathcal{J} = S_{R_1} \cap S_{R_2}$), filtering $R_2$ by tuples that join with $t$ ($\sigma_{\mathcal{J} = \pi_\mathcal{J}(t)}(R_2)$), and recursively apply sampling based on the conditional probability $P(S_{R_2} - \mathcal{J}|\mathcal{J}) = P(S_{R_2})/P(\mathcal{J})$. During sampling, the expensive queries are $P(S_{R_1}),...P(S_{R_n})$, which can be accelerated through caching messages (\Cref{sec:factorizedtreeback}).}

\stitle{Minor Optimizations.} First, we coalesce the messages for \cnt query with those for the tree criterion. For instance, the $c$ element in Variance Semi-ring captures the \cnt statistics. Second, for snowflake schemas where the fact table has N-to-1 relationships with the rest of the tables, we sample the fact table directly~\cite{viswanath1998join}.

\subsubsection{Inter-query Parallelism.}
Parallelism is widely used in ML libraries like \lgbm, which implements parallelized sorting, aggregation, residual updates, and split candidate evaluations. 
For \sys, most DBMSes offer intra-query parallelism, but there can be diminishing returns for individual queries or operations. Thus, \sys also aggressively parallelizes across trees, leaf nodes, candidate splits, and messages.
However, there are dependencies between these queries: a message's query relies on upstream messages, absorption depends on incoming messages, tree nodes depend on ancestor node queries, and gradient boosting trees rely on preceding trees.
To handle this, \sys employs a simple scheduler. Each query $Q$ tracks its dependent queries, and when $Q$ finishes, it sets the ready bit for these dependencies. If all dependencies for a query are set, it's added to a FIFO run queue. Empirically, 4 threads work best for intra-query parallelism, while the rest are used for inter-query parallelism. This reduces gradient boosting training time by $28\%$ and random forest by $35\%$ (\Cref{sec:tpcexp}).

\section{Experiments}

We focus on the fastest alternative: ML libraries (\xgb, \lgbm, \texttt{Sklearn}).  We start with a single-node setting, and then evaluate scalability to the number of features, types of joins, and database size on multiple nodes. 
Finally, we compare with factorized (\lmfao) and non-factorized (\mad) in-DB ML techniques.

\begin{figure}
  \centering
  \includegraphics[width=.45\columnwidth]{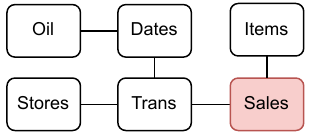}
  \vspace*{-3mm}
  \caption{Favorita schema. \texttt{\red{Sales}} is the fact table.}
  \vspace*{-6mm}
   \label{fav_schema}
\end{figure}

\stitle{Datasets.} We primarily report results using the Favorita~\cite{favorita} dataset used in prior factorized ML work~\cite{schleich2016learning,schleich2019layered} (\Cref{fav_schema}). \texttt{Sales} is the fact table ($2.7GB$, 80M rows), and has N-to-1 relationships with the other dimensions (${<}2MB$ each).  There are 13 features.  We report TPC-DS results for scalability experiments and use IMDB for galaxy schema experiments; additional TPC-H/DS results are in the \Cref{sec:tpcexp}.

\stitle{Preprocess.}   
Although Favorita and TPC-H/DS are standard benchmarks in prior factorized learning evaluations~\cite{schleich2019layered}, their features are  non-predictive and lead to highly unbalanced trees. This  artificially favors \sys since all but one leaf node in the decision tree would contain very few records, and thus take negligible training time---the performance differences with other systems are fully dominated by join materialization.  To ensure balanced trees and fair comparison, we impute one feature attribute in each of the 5 dimension tables with random integers drawn from $[1,1000]$.  Then we impute the target variable as sum of transformed features\footnote{Favorita applies: $y=f_{item}log(f_{items})+log(f_{oil}) - 10f_{dates}-10f_{stores} + f_{trans}^2$}.
Finally, 
we  dictionary encode strings into 32-bit unsigned integers~\cite{aberger2017emptyheaded,raman2013db2} to avoid parsing errors in ML libraries like \lgbm.

\stitle{Models.} We evaluate decision tree, random forest, and gradient boosting.  \sys is intended to complement other DBMS workloads, so all experiments start with data persistent on disk but not in memory. We assume  by default that data are already persisted in the disk-based DBMSes (\dbmsx and disk-based \duckdb). We report the end-to-end training time for decision tree of max depth $10$, and vary the number of trees (iterations) in the random forest and gradient boosting. For \sys, the main cost is from DBMSes, and the Python codes introduce negligible (${<}0.1s$) overhead.

\stitle{Methods.}
We evaluate \sys with different DBMS backends. \dbmsx (\texttt{X-col} and \texttt{X-row} for column and row-oriented storage and execution engines) and \duckdb-disk (\texttt{D-disk}) are disk-based and directly execute queries on the base DBMS, whereas memory-based \duckdb (\texttt{D-mem}) first loads the DBMS from disk.  \texttt{DP} refers to the disk-based \duckdb backend using \pd updates through the \duckdb's relational API, and \texttt{D-Swap} refers to the modified memory-based \duckdb for efficient residual updates (\Cref{sec:duckpandas}).
By default, we use \texttt{D-Swap} as the backend as it has the best performance (\Cref{sec:db_exp}). ML libraries\footnote{Python version \lgbm and \xgb have memory issues~\cite{lightgbmmemory} so we use CLI version.} (\lgbm, \xgb, \texttt{Sklearn}) expect a single CSV as input, so incur the cost to materialize and export the join result (${\sim}7GB$ for Favorita), load the CSV, and train the model.  \duckdb joins and exports the data faster than \dbmsx, so we report its numbers.  

\stitle{Hardware:} We use Azure VM: 16 cores, 128 GB memory, and SSD.

\subsection{Comparison With ML Libraries}
\label{exp:ml}
We compare with SOTA ML libraries for training tree-based models (\lgbm~\cite{ke2017lightgbm}, \xgb~\cite{chen2016xgboost} and \texttt{Sklearn}~\cite{scikit-learn}).  \texttt{Sklearn} implements the standard and histogram-based gradient boosting with algorithms similar to \lgbm, so we report both implementations. We set the number of bins to $1000$ for \lgbm and \xgb, and $255$ for \texttt{Sklearn} (its limit). By default, we train gradient boosting and random forest with best-first growth, with a maximum of 8 leaves per tree. The gradient boosting learning rate is $0.1$. The random forest data sampling rate without replacement is $10\%$, \revise{and feature sampling rate is $80\%$}. We report up to $100$ iterations. The $0^{th}$ iteration reports the join materialization, export, and load costs.

\begin{figure}
\centering
 \begin{subfigure}[t]{0.5\textwidth}
     \centering
     \includegraphics[width=\textwidth]{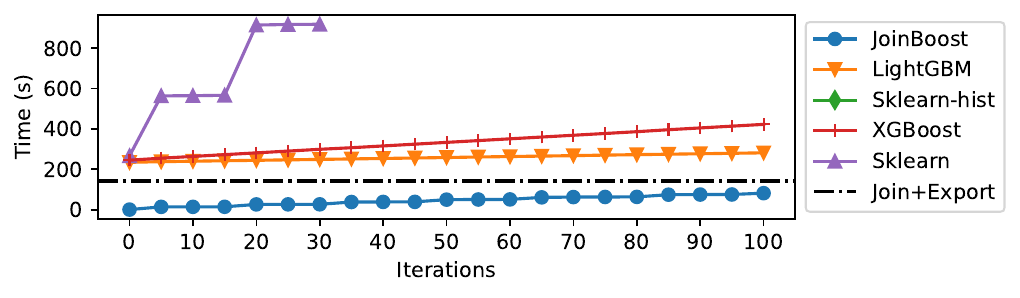}
     \vspace*{-6mm}
      \caption{\revise{Random forest training time. \sys is ${\sim}3\times$ faster than \lgbm.}}
      \label{fig:rf_time}
 \end{subfigure}
 \hfill
  \begin{subfigure}[t]{0.5\textwidth}
     \includegraphics[width=0.79\textwidth]{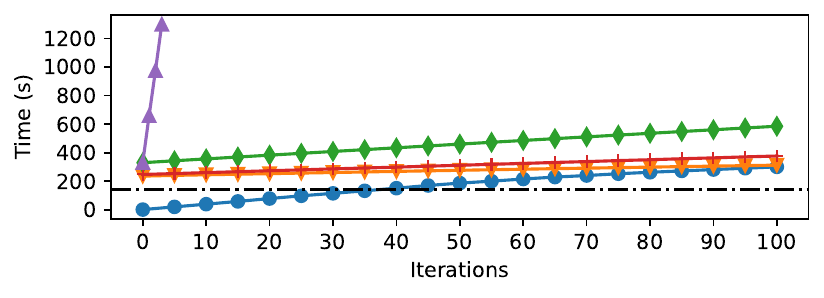}
     \vspace*{-3mm}
      \caption{Gradient boosting Training time. \sys is ${\sim}1.1\times$ faster.}
      \label{fig:gb_time}
 \end{subfigure}
 \hfill
     \begin{subfigure}[t]{0.5\textwidth}
     \includegraphics[width=0.79\textwidth]{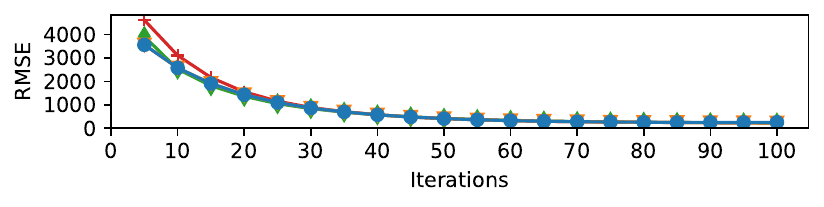}
     \vspace*{-3mm}
      \caption{Gradient boosting Accuracy. The final $rmse$ is nearly identical.}
      \label{fig:gb_acc}
 \end{subfigure}
 \vspace*{-4mm}
\caption{Gradient boosting and random forest training time and accuracy on Favorita compared to ML libraries when the materialized join fits into memory on a single node. 
} 
\vspace*{-5mm}
\end{figure}

\Cref{fig:rf_time} shows random forest results. \sys is ${\sim}3\times$ faster than \lgbm by avoiding materialization and export costs (dotted black line), and loading costs; it also parallelizes across trees.  In fact, \sys finishes 100 iterations before the export is done.   \texttt{Sklearn} also parallelizes across trees, but is so slow that we terminate after 32 iterations. The final model error ($rmse$) is nearly identical (${\sim}2350$) for \sys, \lgbm and \xgb.

\Cref{fig:gb_time} shows gradient boosting results. \sys is ${\sim}1.1\times$ faster than \lgbm, and is ${\sim}1.2\times$ faster than \xgb by avoiding materialization and export costs.  
\Cref{fig:gb_acc} shows the model $rmse$. \sys and \lgbm have equivalent $rmse$ as both employ the same algorithm, while the final $rmse$ is similar across all. 
Note that the models begin to converge $\sim60$ iterations; \sys converges by the time \lgbm and \xgb have loaded their data.

\begin{figure}
  \includegraphics[width=0.8\columnwidth]{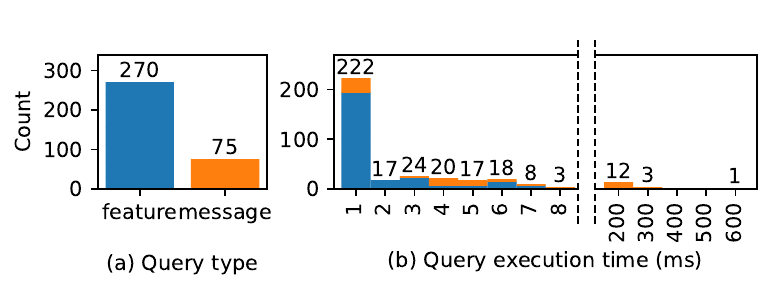}
  \vspace*{-5mm}
      \caption{
     \revise{The $1^{st}$ iteration of \sys gradient boosting. (a) The number of queries for passing messages and finding the best split. (b) The histogram of query execution time (ms).}}
     \vspace*{-3mm}
      \label{fig:query_details}
\end{figure}

\revise{To interpret the \sys cost, we zoom in on the  $1^{st}$ iteration of \sys gradient boosting in \Cref{fig:query_details}, which displays the total number of queries for passing messages (orange) and finding the best split (blue) along with a histogram for the query execution time distribution. With a tree of $8$ leaves and $15$ nodes, there are $270 = 15\times 18$ (number of features) queries for feature split, and $75 = 15\times 5$ (number of join edges) queries for message passing, as expected. All feature split queries are efficient, taking $<10ms$.
The performance bottlenecks are the queries passing messages from the fact table, which require join-aggregation, result materialization, and take $>200ms$. 
This highlights the importance of aggressively reusing messages across nodes (\Cref{alg:factorizedtree}).}

\begin{figure}
     \centering
  \includegraphics[width=0.75\columnwidth]{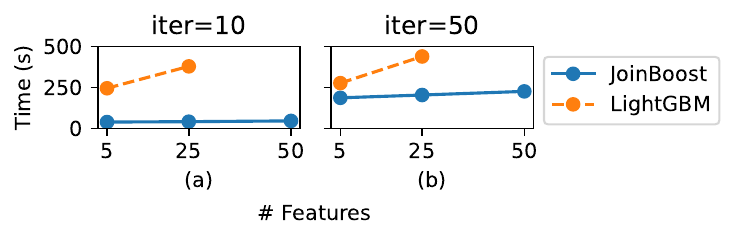}
  \vspace*{-6mm}
      \caption{
      Gradient boosting training time of $10^{th}$ (a) and $50^{th}$ (b) iteration when varying $\#$ of imputed features (x-axis). \lgbm runs out of memory when imputing 50 features.}
      \vspace*{-3mm}
      \label{fig:scalafeature}
\end{figure}
\begin{figure}
  \includegraphics[width=0.77\columnwidth]{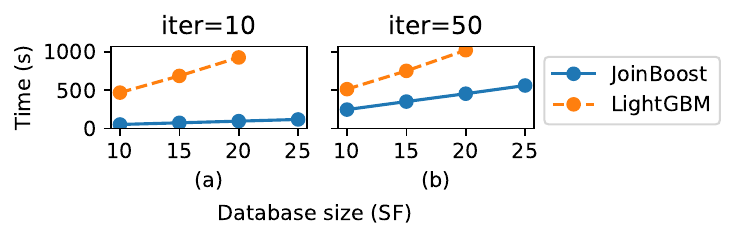}
  \vspace*{-5mm}
      \caption{
     Gradient boosting training time of $10^{th}$ (a) and $50^{th}$ (b) iteration.  The X-axis varies TPC-DS SF (database size). \lgbm runs out of memory when SF=25.}
      \label{fig:scaladata}
\end{figure}
\begin{figure}
  \includegraphics[width=0.8\columnwidth]{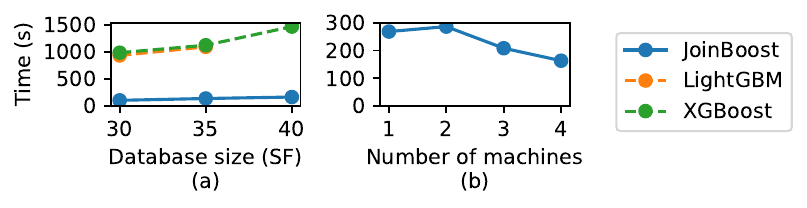}
  \vspace*{-5mm}
      \caption{
     \revise{Gradient boosting training time of $10$ iterations  (a) varying SF on 4 machines (b) varying $\#$ of machines for SF=40. \lgbm runs out of memory at SF=40 even on 4 machines.}}
     \vspace*{-6mm}
      \label{fig:scaladata2}
\end{figure}

\subsection{Scalability}
\label{sec:scala}
We now study scalability to \# features, DB size, and join complexity.  
\stitle{\# Features.}  We train gradient boosting using Favorita and vary the number of features from $5$ to $50$, and report training time at the $10^{th}$ and $50^{th}$ iterations. \Cref{fig:scalafeature} shows that \lgbm slows by ${>}1.5\times$ with $25$ features, and runs out of memory (125GB) at $50$ features.   \xgb supports out-of-core training, but  took${\sim}4000s$ to train $50$ features for 10 iterations, and wrote ${\sim}80GB$ of intermediate results to disk (not plotted). \sys scales linearly with ${>}10\times$ lower slope, and scales to out-of-core execution thanks to the DBMS.

\stitle{Single-node Scalability.} Favorita is a fixed dataset, so we use TPC-DS (145 features) to scale the database ($SF\in[10,25]$).   \Cref{fig:scaladata} shows that both systems scale linearly, but \sys has a lower slope (${\sim}10\times$ lower at $10^{th}$ iteration, and ${\sim}2\times$ lower at $50^{th}$). \lgbm runs out of memory at $SF=25$.

\stitle{Multi-node Scalability.} 
\revise{We use the Dask version~\cite{rocklin2015dask} of \lgbm and \xgb, and Dask-SQL for \sys to train gradient boosting  on multiple machines for 10 iterations on TPC-DS $SF{\in}[30,40]$. We use 4 n1-highmem-16 GCP instances (16 vCPUs, 104 GB RAM each) with data replicated across all instances; at least 4 machines are needed for \lgbm to run on $SF=30$ due to large join sizes and inefficient memory usage~\cite{lightgbmmemory}. \Cref{fig:scaladata2} (a) shows that, on 4 machines, all systems scale linearly , but \sys is ${>}9\times$ faster with a ${\sim}5\times$ lower slope. For $SF{=}40$, \lgbm runs out of memory even on 4 machines.  In contrast, \Cref{fig:scaladata2} (b) shows that for $SF{=}40$, \sys can train on a single machine, outperforms \xgb (using 4 machines), and speeds up with more machines.}

\begin{figure}
     \centering
  \includegraphics[width=0.35\columnwidth]{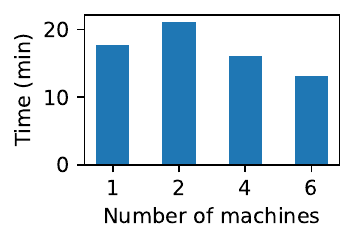}
  \vspace*{-5mm}
      \caption{
      Decision Tree Training Time over TPC-DS (SF=1000) in Data Warehouse. Increasing the number of machines to 4 (6) reduces the training time by $10\%$ ($25\%$).}
      \vspace*{-5mm}
      \label{fig:datawarehouse}
\end{figure}

\stitle{Cloud-warehouse Scalability.} We use a cloud data warehouse \texttt{DW-X} to train a decision tree with max depth 3 over TPC-DS SF=1000, and study multi-machine scalability.    Each machine has 74 cores, 300GB of memory, and SSDs.  
 We replicate dimensional tables across the machines, and hash-partition the fact table.   \Cref{fig:datawarehouse} shows that 2 machines introduce a shuffle stage that slows training, and increasing to 4 (6) machines reduces training by $10\%$ ($25\%$).

\begin{figure}
\centering
     \centering
     \includegraphics[width=0.17\textwidth]{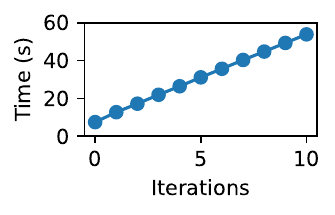} 
     \vspace*{-5mm}
\caption{Gradient Boosting Training over IMDB dataset with a galaxy schema. Each tree takes ${\sim}5s$.  ML libraries do not run because the join is too large to materialize.}
\vspace*{-6mm}
\label{fig:imdbtime}
\end{figure}

\stitle{Galaxy Schemas.}
Galaxy schemas have N-to-N relationships that are prohibitive to materialize.   We use clustered predicate trees (\Cref{sec:predtree}) to train gradient boosting on the IMDB dataset (\Cref{fig:cluster}).  \texttt{Cast\_Info} is ${\sim}1GB$ and the total DB is $1.2GB$.   \sys scales linearly with the number of iterations (\Cref{fig:imdbtime}); it trains one tree and updates residuals in each cluster's fact table within ${\sim}5s$. \lgbm  cannot run because the join result is ${>}1TB$.

\subsection{Effect of DBMSes}
\label{sec:db_exp}

We now use Favorita to train 1 iteration of gradient boosting, and compare train and update costs for different DBMSes (\texttt{DBMS-X} and \duckdb). 
To study the potential benefits of column swap (\Cref{sec:duckpandas}) in commercial DBMSes, \texttt{X-Swap*} reports the theoretical cost using the time to create the updated column as a new table (since column swap is then ``free'').
\begin{figure}
\centering
\includegraphics[width=0.35\textwidth]{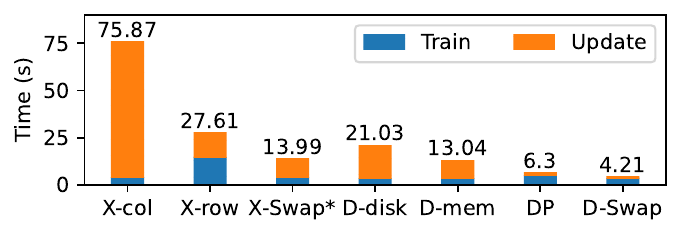}
\vspace*{-6mm}
\caption{Training and residual update time for one decision tree on Favorita for different DBMSes. \texttt{X-Swap*} is the theoretical runtime by simulating column swaps.}
\vspace*{-5mm}
\label{fig:breakdown}
\end{figure}

\Cref{fig:breakdown} breaks down train and update costs.
Decision trees and random forests only require training (blue bar), which is dominated by columnar execution: \texttt{X-col} and \duckdb take $3.2-3.9s$ versus $14.5s$ for \texttt{X-row}. 
Gradient boosting introduces high update costs across all of the baseline DBMSes.  We see that using Pandas to perform the update (\texttt{DP}) reduces residual updates by ${\sim}15\times$ (${17.8s}{\to}{1.2s}$), but slows training by $60\%$ (${3.2s}{\to}{5.1s}$) due to \duckdb-\pd interop overhead. \texttt{D-opt} implements column swapping inside \duckdb and improves training. We also see that adding column swapping to \texttt{DBMS-X} (\texttt{X-Swap*}) leads to respectable gradient boosting performance (${\sim}3\times$ of \texttt{D-Swap}) but can benefit from its out-of-memory and multi-node scalability features.

\vspace{-2mm}
\subsection{Comparison With In-DB ML Techniques}
\label{exp:lmfao}
\vspace{-1mm}
We first compare with the dominant factorized approach (\lmfao~\cite{schleich2019layered}), followed by the non-factorized approach (\mad).

\stitle{Factorized ML.}
We first compare with \lmfao~\cite{schleich2019layered}, which supports decision trees (but not gradient boosting or random forests).  We train a decision tree (max depth=10) with best-first growth; the trained tree is balanced and has 1024 leaves.
Via correspondence, the \lmfao authors shared a version that compiles a program for the queries used to split the root tree node, and reuses it for growing the rest of the tree.  
We set the highest optimization level for \lmfao, and exclude the time for query compilation (${\sim} 15s$) and data loading. 
To separate algorithmic vs implementation differences, we implement two variations of \sys.   \texttt{Naive} materializes the join result without factorized ML.   \texttt{Batch} implements \lmfao's core logical optimizations (Multi Root, Aggregate Push-down and Merge View) for decision tree; this corresponds to message passing with message re-use within, but not between tree nodes. 

\begin{figure}
\centering
  \begin{subfigure}[t]{0.24\textwidth}
     \centering
     \includegraphics[width=\textwidth]{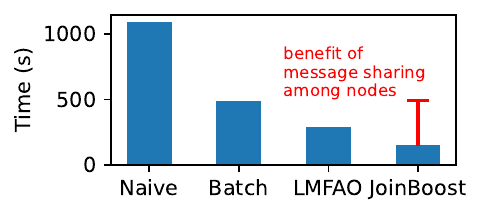}
     \vspace*{-7mm}
     \caption{}
     \label{fig:algdiff}
 \end{subfigure}
   \begin{subfigure}[t]{0.11\textwidth}
     \centering
     \includegraphics[width=\textwidth]{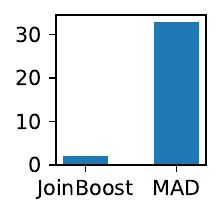}
      \vspace*{-7mm}
     \caption{}
     \label{fig:indbml}
 \end{subfigure}
 \vspace*{-6mm}
\caption{(a) Training time for decision tree. \sys's caching and work sharing improve over \lmfao, naive materialization and query batching variants. (c) Training time for decision tree. \sys is ${\sim}16\times$ faster than \mad.}
\vspace*{-5mm}
\end{figure}

\Cref{fig:algdiff} reports the training time. Even  with the custom engine and specialized optimizations, \lmfao is still ${\sim}1.9\times$ slower than \sys due to the lack of message caching (\Cref{alg:factorizedtree}).  By eliminating the implementation differences, \texttt{Batch} demonstrates that message caching improves training by ${\sim}3\times$. The improvement is because half the messages across tree nodes are cached, and the sizes of the cached messages tend to be much larger; the intuition is that the messages outgoing from the relation containing the split attribute will be smaller, since the split predicate has been applied. In theory, two-pass semi-join reduction~\cite{abo2016faq,olteanu2015size} could reduce message sizes, but the high overheads outweigh benefits~\cite{aberger2017emptyheaded}. \texttt{Batch} is ${\sim}2\times$ faster than \texttt{Naive} due to factorization and shared work.

\stitle{Non-Factorized ML.}
\mad~\cite{hellerstein2012madlib} is a PostgreSQL extension that supports ML using user-defined types and functions.  \mad doesn't apply factorized ML, so the join has to be materialized.  \mad times out after 1 hour when training a decision tree model (max depth=10) on the full Favorita dataset, so we reduced the training data size to 10k rows; for \sys, we reduced the fact table to 10k rows.   \Cref{fig:indbml} shows that \sys is ${\sim}16\times$ faster than \mad. This reproduces prior findings~\cite{schleich2019layered}, and is likely due to the lack of factorized ML and an inefficient implementation.

\vspace{-3mm}
\section{Limitations and Future Work}
\label{sec:limitation}
\vspace{-1mm}
\stitle{Limitations.}
\revise{\sys has some limitations that may prevent its widespread use. For usability, it requires users (e.g., data scientists) to specify schema and join conditions if they cannot be inferred automatically. In terms of algorithm, \sys currently supports the most widely used criterion, $rmse$, for galaxy schema using {\it Clustered Predicate Trees} (CPT). However, efficiently factorizing residual updates for other criteria remains challenging (\Cref{sec:boosting}).}

\stitle{Future work.}
\revise{In the future, our goals include enhancing \sys's usability by integrating it with "semantic layers"\cite{chatziantoniou2020data,dbtdatamodel,lookerdatamodel,tableaudatamodel} that transform raw data into tables with meaningful attributes for domain scientists and extending them for join graphs. To support more criteria, 
one promising future direction is to apply polynomial approximations~\cite{huggins2017pass}: variance semi-ring computes $2^{nd}$ order monomials $(1,y,y^2)$, which can be extended to higher order $(1, y, y^2,..., y^k)$ with tailored multiplication operation to both   update monomials and preserve addition-to-multiplication property.
Monomials can approximate various criteria by Taylor expansion.}

\revise{Additionally, applying \sys to GPU-based DBMSes~\cite{joinboostgpu} is of interest, as GPU memory size is notoriously limited~\cite{asada2022share} especially for join, and \sys may address this. We also aim to explore data augmentation search~\cite{chepurko2020arda}, which joins training data with additional features and  evaluates model (random forest) improvement. \sys's scalability and portability can be leveraged for more efficient augmentation search in enterprise data warehouses. Lastly, examining the explainability of gradient boosting, particularly recent developments in causal inference~\cite{bakhitov2022causal} with potential interactions between causal and join graphs, will offer insights into gradient boosting  decision-making processes.}

\vspace{-3mm}
\section{Conclusion}
\vspace{-1mm}
\sys is the first In-DB factorized ML system for tree models (decision trees, random forests, and gradient boosting) with only SQL.  \sys is comparable or faster than the SOTA \lgbm ML library on in-memory datasets, but scales well beyond \lgbm's capabilities in terms of \# of features, database size, and join graph complexity.  \sys exposes a Python API that mimics \lgbm's API and is portable to any DBMS and dataframe library.

\pagebreak

\bibliographystyle{ACM-Reference-Format}
\bibliography{main}


\begin{thebibliography}{80}


\ifx \showCODEN    \undefined \def \showCODEN     #1{\unskip}     \fi
\ifx \showDOI      \undefined \def \showDOI       #1{#1}\fi
\ifx \showISBNx    \undefined \def \showISBNx     #1{\unskip}     \fi
\ifx \showISBNxiii \undefined \def \showISBNxiii  #1{\unskip}     \fi
\ifx \showISSN     \undefined \def \showISSN      #1{\unskip}     \fi
\ifx \showLCCN     \undefined \def \showLCCN      #1{\unskip}     \fi
\ifx \shownote     \undefined \def \shownote      #1{#1}          \fi
\ifx \showarticletitle \undefined \def \showarticletitle #1{#1}   \fi
\ifx \showURL      \undefined \def \showURL       {\relax}        \fi
\providecommand\bibfield[2]{#2}
\providecommand\bibinfo[2]{#2}
\providecommand\natexlab[1]{#1}
\providecommand\showeprint[2][]{arXiv:#2}

\bibitem[azu({[n.\,d.]})]%
        {azureML}
 \bibinfo{year}{[n.\,d.]}\natexlab{}.
\newblock \bibinfo{title}{Azure Machine Learning documentation}.
\newblock
  \bibinfo{howpublished}{\url{https://learn.microsoft.com/en-us/azure/machine-learning/}}.
\newblock


\bibitem[PDP({[n.\,d.]})]%
        {PDPA}
 \bibinfo{year}{[n.\,d.]}\natexlab{}.
\newblock \bibinfo{title}{Personal-Data-Protection-Act}.
\newblock
  \bibinfo{howpublished}{\url{https://www.pdpc.gov.sg/Overview-of-PDPA/The-Legislation/Personal-Data-Protection-Act}}.
\newblock


\bibitem[sno({[n.\,d.]})]%
        {snowflakeML}
 \bibinfo{year}{[n.\,d.]}\natexlab{}.
\newblock \bibinfo{title}{Snowflake Machine Learning Platforms}.
\newblock
  \bibinfo{howpublished}{\url{https://www.snowflake.com/guides/machine-learning-platforms}}.
\newblock


\bibitem[red({[n.\,d.]})]%
        {redshiftML}
 \bibinfo{year}{[n.\,d.]}\natexlab{}.
\newblock \bibinfo{title}{Using machine learning in Amazon Redshift}.
\newblock
  \bibinfo{howpublished}{\url{https://docs.amazonaws.cn/en_us/redshift/latest/dg/machine_learning.html}}.
\newblock


\bibitem[kot(2006)]%
        {kotidis2006updates}
 \bibinfo{year}{2006}\natexlab{}.
\newblock \showarticletitle{Updates through views: A new hope}. In
  \bibinfo{booktitle}{\emph{22nd International Conference on Data Engineering
  (ICDE'06)}}. IEEE, \bibinfo{pages}{2--2}.
\newblock


\bibitem[imd(2013)]%
        {imdb}
 \bibinfo{year}{2013}\natexlab{}.
\newblock \bibinfo{title}{IMDB}.
\newblock \bibinfo{howpublished}{\url{https://www.imdb.com/interfaces/}}.
\newblock


\bibitem[fav(2017)]%
        {favorita}
 \bibinfo{year}{2017}\natexlab{}.
\newblock \bibinfo{title}{Corporación Favorita Grocery Sales Forecasting}.
\newblock
  \bibinfo{howpublished}{\url{https://www.kaggle.com/c/favorita-grocery-sales-forecasting}}.
\newblock


\bibitem[lig(2017)]%
        {lightgbmmemory}
 \bibinfo{year}{2017}\natexlab{}.
\newblock \bibinfo{title}{Lightgbm memory explodes in start train}.
\newblock
  \bibinfo{howpublished}{\url{https://github.com/microsoft/LightGBM/issues/1032}}.
\newblock


\bibitem[loo(2020)]%
        {lookerdatamodel}
 \bibinfo{year}{2020}\natexlab{}.
\newblock \bibinfo{title}{Looker data modeling}.
\newblock
  \bibinfo{howpublished}{\url{https://www.looker.com/platform/data-modeling/}}.
\newblock


\bibitem[tab(2020)]%
        {tableaudatamodel}
 \bibinfo{year}{2020}\natexlab{}.
\newblock \bibinfo{title}{The Tableau Data Model}.
\newblock
  \bibinfo{howpublished}{\url{https://help.tableau.com/current/online/en-us/datasource_datamodel.htm}}.
\newblock


\bibitem[kag(2021)]%
        {kaggle_survey}
 \bibinfo{year}{2021}\natexlab{}.
\newblock \bibinfo{title}{2021 Kaggle Data Science and Machine Learning
  Survey}.
\newblock
  \bibinfo{howpublished}{\url{https://www.kaggle.com/code/paultimothymooney/2021-kaggle-data-science-machine-learning-survey/notebook}}.
\newblock


\bibitem[rel(2021)]%
        {relationalapi}
 \bibinfo{year}{2021}\natexlab{}.
\newblock \bibinfo{title}{Client APIs Overview}.
\newblock \bibinfo{howpublished}{\url{https://duckdb.org/docs/api/overview}}.
\newblock


\bibitem[tec(2021)]%
        {tech}
 \bibinfo{year}{2021}\natexlab{}.
\newblock \bibinfo{title}{(Technical Report) JoinBoost: In-Database Tree-Models
  In Action}.
\newblock
  \bibinfo{howpublished}{\url{https://anonymous.4open.science/r/JoinBoost-FBC4/technical/JoinBoost_tech.pdf}}.
\newblock


\bibitem[Cli(2022)]%
        {ClickBench}
 \bibinfo{year}{2022}\natexlab{}.
\newblock \bibinfo{title}{ClickBench: a Benchmark For Analytical Databases}.
\newblock \bibinfo{howpublished}{\url{https://benchmark.clickhouse.com/}}.
\newblock


\bibitem[dbt(2023)]%
        {dbtdatamodel}
 \bibinfo{year}{2023}\natexlab{}.
\newblock \bibinfo{title}{The dbt Semantic Layer}.
\newblock
  \bibinfo{howpublished}{\url{https://www.getdbt.com/product/semantic-layer/}}.
\newblock


\bibitem[Abadi et~al\mbox{.}(2015)]%
        {tensorflow2015-whitepaper}
\bibfield{author}{\bibinfo{person}{Mart\'{i}n Abadi}, \bibinfo{person}{Ashish
  Agarwal}, \bibinfo{person}{Paul Barham}, \bibinfo{person}{Eugene Brevdo},
  \bibinfo{person}{Zhifeng Chen}, \bibinfo{person}{Craig Citro},
  \bibinfo{person}{Greg~S. Corrado}, \bibinfo{person}{Andy Davis},
  \bibinfo{person}{Jeffrey Dean}, \bibinfo{person}{Matthieu Devin},
  \bibinfo{person}{Sanjay Ghemawat}, \bibinfo{person}{Ian Goodfellow},
  \bibinfo{person}{Andrew Harp}, \bibinfo{person}{Geoffrey Irving},
  \bibinfo{person}{Michael Isard}, \bibinfo{person}{Yangqing Jia},
  \bibinfo{person}{Rafal Jozefowicz}, \bibinfo{person}{Lukasz Kaiser},
  \bibinfo{person}{Manjunath Kudlur}, \bibinfo{person}{Josh Levenberg},
  \bibinfo{person}{Dandelion Man\'{e}}, \bibinfo{person}{Rajat Monga},
  \bibinfo{person}{Sherry Moore}, \bibinfo{person}{Derek Murray},
  \bibinfo{person}{Chris Olah}, \bibinfo{person}{Mike Schuster},
  \bibinfo{person}{Jonathon Shlens}, \bibinfo{person}{Benoit Steiner},
  \bibinfo{person}{Ilya Sutskever}, \bibinfo{person}{Kunal Talwar},
  \bibinfo{person}{Paul Tucker}, \bibinfo{person}{Vincent Vanhoucke},
  \bibinfo{person}{Vijay Vasudevan}, \bibinfo{person}{Fernanda Vi\'{e}gas},
  \bibinfo{person}{Oriol Vinyals}, \bibinfo{person}{Pete Warden},
  \bibinfo{person}{Martin Wattenberg}, \bibinfo{person}{Martin Wicke},
  \bibinfo{person}{Yuan Yu}, {and} \bibinfo{person}{Xiaoqiang Zheng}.}
  \bibinfo{year}{2015}\natexlab{}.
\newblock \bibinfo{title}{{TensorFlow}: Large-Scale Machine Learning on
  Heterogeneous Systems}.
\newblock
\newblock
\urldef\tempurl%
\url{https://www.tensorflow.org/}
\showURL{%
\tempurl}
\newblock
\shownote{Software available from tensorflow.org}.


\bibitem[Aberger et~al\mbox{.}(2017)]%
        {aberger2017emptyheaded}
\bibfield{author}{\bibinfo{person}{Christopher~R Aberger},
  \bibinfo{person}{Andrew Lamb}, \bibinfo{person}{Susan Tu},
  \bibinfo{person}{Andres N{\"o}tzli}, \bibinfo{person}{Kunle Olukotun}, {and}
  \bibinfo{person}{Christopher R{\'e}}.} \bibinfo{year}{2017}\natexlab{}.
\newblock \showarticletitle{Emptyheaded: A relational engine for graph
  processing}.
\newblock \bibinfo{journal}{\emph{ACM Transactions on Database Systems (TODS)}}
  \bibinfo{volume}{42}, \bibinfo{number}{4} (\bibinfo{year}{2017}),
  \bibinfo{pages}{1--44}.
\newblock


\bibitem[Abo~Khamis et~al\mbox{.}(2016)]%
        {abo2016faq}
\bibfield{author}{\bibinfo{person}{Mahmoud Abo~Khamis}, \bibinfo{person}{Hung~Q
  Ngo}, {and} \bibinfo{person}{Atri Rudra}.} \bibinfo{year}{2016}\natexlab{}.
\newblock \showarticletitle{FAQ: questions asked frequently}. In
  \bibinfo{booktitle}{\emph{Proceedings of the 35th ACM SIGMOD-SIGACT-SIGAI
  Symposium on Principles of Database Systems}}. \bibinfo{pages}{13--28}.
\newblock


\bibitem[Anghel et~al\mbox{.}(2018)]%
        {anghel2018benchmarking}
\bibfield{author}{\bibinfo{person}{Andreea Anghel}, \bibinfo{person}{Nikolaos
  Papandreou}, \bibinfo{person}{Thomas Parnell}, \bibinfo{person}{Alessandro
  De~Palma}, {and} \bibinfo{person}{Haralampos Pozidis}.}
  \bibinfo{year}{2018}\natexlab{}.
\newblock \showarticletitle{Benchmarking and optimization of gradient boosting
  decision tree algorithms}.
\newblock \bibinfo{journal}{\emph{arXiv preprint arXiv:1809.04559}}
  (\bibinfo{year}{2018}).
\newblock


\bibitem[Asada et~al\mbox{.}(2022)]%
        {asada2022share}
\bibfield{author}{\bibinfo{person}{Yuki Asada}, \bibinfo{person}{Victor Fu},
  \bibinfo{person}{Apurva Gandhi}, \bibinfo{person}{Advitya Gemawat},
  \bibinfo{person}{Lihao Zhang}, \bibinfo{person}{Dong He},
  \bibinfo{person}{Vivek Gupta}, \bibinfo{person}{Ehi Nosakhare},
  \bibinfo{person}{Dalitso Banda}, \bibinfo{person}{Rathijit Sen},
  {et~al\mbox{.}}} \bibinfo{year}{2022}\natexlab{}.
\newblock \showarticletitle{Share the tensor tea: how databases can leverage
  the machine learning ecosystem}.
\newblock \bibinfo{journal}{\emph{arXiv preprint arXiv:2209.04579}}
  (\bibinfo{year}{2022}).
\newblock


\bibitem[Bakhitov and Singh(2022)]%
        {bakhitov2022causal}
\bibfield{author}{\bibinfo{person}{Edvard Bakhitov} {and}
  \bibinfo{person}{Amandeep Singh}.} \bibinfo{year}{2022}\natexlab{}.
\newblock \showarticletitle{Causal gradient boosting: Boosted instrumental
  variable regression}. In \bibinfo{booktitle}{\emph{Proceedings of the 23rd
  ACM Conference on Economics and Computation}}. \bibinfo{pages}{604--605}.
\newblock


\bibitem[Boehm et~al\mbox{.}(2016)]%
        {boehm2016systemml}
\bibfield{author}{\bibinfo{person}{Matthias Boehm}, \bibinfo{person}{Michael~W
  Dusenberry}, \bibinfo{person}{Deron Eriksson}, \bibinfo{person}{Alexandre~V
  Evfimievski}, \bibinfo{person}{Faraz~Makari Manshadi},
  \bibinfo{person}{Niketan Pansare}, \bibinfo{person}{Berthold Reinwald},
  \bibinfo{person}{Frederick~R Reiss}, \bibinfo{person}{Prithviraj Sen},
  \bibinfo{person}{Arvind~C Surve}, {et~al\mbox{.}}}
  \bibinfo{year}{2016}\natexlab{}.
\newblock \showarticletitle{Systemml: Declarative machine learning on spark}.
\newblock \bibinfo{journal}{\emph{Proceedings of the VLDB Endowment}}
  \bibinfo{volume}{9}, \bibinfo{number}{13} (\bibinfo{year}{2016}),
  \bibinfo{pages}{1425--1436}.
\newblock


\bibitem[Breiman(2001)]%
        {breiman2001random}
\bibfield{author}{\bibinfo{person}{Leo Breiman}.}
  \bibinfo{year}{2001}\natexlab{}.
\newblock \showarticletitle{Random forests}.
\newblock \bibinfo{journal}{\emph{Machine learning}} \bibinfo{volume}{45},
  \bibinfo{number}{1} (\bibinfo{year}{2001}), \bibinfo{pages}{5--32}.
\newblock


\bibitem[Breiman et~al\mbox{.}(2017)]%
        {breiman2017classification}
\bibfield{author}{\bibinfo{person}{Leo Breiman}, \bibinfo{person}{Jerome~H
  Friedman}, \bibinfo{person}{Richard~A Olshen}, {and}
  \bibinfo{person}{Charles~J Stone}.} \bibinfo{year}{2017}\natexlab{}.
\newblock \bibinfo{booktitle}{\emph{Classification and regression trees}}.
\newblock \bibinfo{publisher}{Routledge}.
\newblock


\bibitem[Ch and Ghane(1993)]%
        {ch1993sampling}
\bibfield{author}{\bibinfo{person}{Bishop~PRML Ch} {and}
  \bibinfo{person}{Alireza Ghane}.} \bibinfo{year}{1993}\natexlab{}.
\newblock \showarticletitle{Sampling Methods}.
\newblock  (\bibinfo{year}{1993}).
\newblock


\bibitem[Chatziantoniou and Kantere(2020)]%
        {chatziantoniou2020data}
\bibfield{author}{\bibinfo{person}{Damianos Chatziantoniou} {and}
  \bibinfo{person}{Verena Kantere}.} \bibinfo{year}{2020}\natexlab{}.
\newblock \showarticletitle{Data Virtual Machines: Data-Driven Conceptual
  Modeling of Big Data Infrastructures.}. In
  \bibinfo{booktitle}{\emph{EDBT/ICDT Workshops}}.
\newblock


\bibitem[Chen et~al\mbox{.}(2016)]%
        {chen2016towards}
\bibfield{author}{\bibinfo{person}{Lingjiao Chen}, \bibinfo{person}{Arun
  Kumar}, \bibinfo{person}{Jeffrey Naughton}, {and} \bibinfo{person}{Jignesh~M
  Patel}.} \bibinfo{year}{2016}\natexlab{}.
\newblock \showarticletitle{Towards linear algebra over normalized data}.
\newblock \bibinfo{journal}{\emph{arXiv preprint arXiv:1612.07448}}
  (\bibinfo{year}{2016}).
\newblock


\bibitem[Chen and Guestrin(2016)]%
        {chen2016xgboost}
\bibfield{author}{\bibinfo{person}{Tianqi Chen} {and} \bibinfo{person}{Carlos
  Guestrin}.} \bibinfo{year}{2016}\natexlab{}.
\newblock \showarticletitle{Xgboost: A scalable tree boosting system}. In
  \bibinfo{booktitle}{\emph{Proceedings of the 22nd acm sigkdd international
  conference on knowledge discovery and data mining}}.
  \bibinfo{pages}{785--794}.
\newblock


\bibitem[Chepurko et~al\mbox{.}(2020)]%
        {chepurko2020arda}
\bibfield{author}{\bibinfo{person}{Nadiia Chepurko}, \bibinfo{person}{Ryan
  Marcus}, \bibinfo{person}{Emanuel Zgraggen}, \bibinfo{person}{Raul~Castro
  Fernandez}, \bibinfo{person}{Tim Kraska}, {and} \bibinfo{person}{David
  Karger}.} \bibinfo{year}{2020}\natexlab{}.
\newblock \showarticletitle{ARDA: automatic relational data augmentation for
  machine learning}.
\newblock \bibinfo{journal}{\emph{arXiv preprint arXiv:2003.09758}}
  (\bibinfo{year}{2020}).
\newblock


\bibitem[Chollet et~al\mbox{.}(2015)]%
        {chollet2015keras}
\bibfield{author}{\bibinfo{person}{Francois Chollet} {et~al\mbox{.}}}
  \bibinfo{year}{2015}\natexlab{}.
\newblock \bibinfo{booktitle}{\emph{Keras}}.
\newblock
\urldef\tempurl%
\url{https://github.com/fchollet/keras}
\showURL{%
\tempurl}


\bibitem[Cunningham et~al\mbox{.}(2004)]%
        {cunningham2004pivot}
\bibfield{author}{\bibinfo{person}{Conor Cunningham},
  \bibinfo{person}{C{\'e}sar~A Galindo-Legaria}, {and} \bibinfo{person}{Goetz
  Graefe}.} \bibinfo{year}{2004}\natexlab{}.
\newblock \showarticletitle{PIVOT and UNPIVOT: Optimization and Execution
  Strategies in an RDBMS}. In \bibinfo{booktitle}{\emph{Proceedings of the
  Thirtieth international conference on Very large data bases-Volume 30}}.
  \bibinfo{pages}{998--1009}.
\newblock


\bibitem[Curtin et~al\mbox{.}(2020)]%
        {curtin2020rk}
\bibfield{author}{\bibinfo{person}{Ryan Curtin}, \bibinfo{person}{Benjamin
  Moseley}, \bibinfo{person}{Hung Ngo}, \bibinfo{person}{XuanLong Nguyen},
  \bibinfo{person}{Dan Olteanu}, {and} \bibinfo{person}{Maximilian Schleich}.}
  \bibinfo{year}{2020}\natexlab{}.
\newblock \showarticletitle{Rk-means: Fast clustering for relational data}. In
  \bibinfo{booktitle}{\emph{International Conference on Artificial Intelligence
  and Statistics}}. PMLR, \bibinfo{pages}{2742--2752}.
\newblock


\bibitem[Feng et~al\mbox{.}(2012)]%
        {feng2012towards}
\bibfield{author}{\bibinfo{person}{Xixuan Feng}, \bibinfo{person}{Arun Kumar},
  \bibinfo{person}{Benjamin Recht}, {and} \bibinfo{person}{Christopher
  R{\'e}}.} \bibinfo{year}{2012}\natexlab{}.
\newblock \showarticletitle{Towards a unified architecture for in-RDBMS
  analytics}. In \bibinfo{booktitle}{\emph{Proceedings of the 2012 ACM SIGMOD
  International Conference on Management of Data}}. \bibinfo{pages}{325--336}.
\newblock


\bibitem[Friedman(2002)]%
        {friedman2002stochastic}
\bibfield{author}{\bibinfo{person}{Jerome~H Friedman}.}
  \bibinfo{year}{2002}\natexlab{}.
\newblock \showarticletitle{Stochastic gradient boosting}.
\newblock \bibinfo{journal}{\emph{Computational statistics \& data analysis}}
  \bibinfo{volume}{38}, \bibinfo{number}{4} (\bibinfo{year}{2002}),
  \bibinfo{pages}{367--378}.
\newblock


\bibitem[Gray et~al\mbox{.}(1997)]%
        {gray1997data}
\bibfield{author}{\bibinfo{person}{Jim Gray}, \bibinfo{person}{Surajit
  Chaudhuri}, \bibinfo{person}{Adam Bosworth}, \bibinfo{person}{Andrew Layman},
  \bibinfo{person}{Don Reichart}, \bibinfo{person}{Murali Venkatrao},
  \bibinfo{person}{Frank Pellow}, {and} \bibinfo{person}{Hamid Pirahesh}.}
  \bibinfo{year}{1997}\natexlab{}.
\newblock \showarticletitle{Data cube: A relational aggregation operator
  generalizing group-by, cross-tab, and sub-totals}.
\newblock \bibinfo{journal}{\emph{Data mining and knowledge discovery}}
  \bibinfo{volume}{1}, \bibinfo{number}{1} (\bibinfo{year}{1997}),
  \bibinfo{pages}{29--53}.
\newblock


\bibitem[Green et~al\mbox{.}(2007)]%
        {green2007provenance}
\bibfield{author}{\bibinfo{person}{Todd~J Green}, \bibinfo{person}{Grigoris
  Karvounarakis}, {and} \bibinfo{person}{Val Tannen}.}
  \bibinfo{year}{2007}\natexlab{}.
\newblock \showarticletitle{Provenance semirings}. In
  \bibinfo{booktitle}{\emph{Proceedings of the twenty-sixth ACM
  SIGMOD-SIGACT-SIGART symposium on Principles of database systems}}.
  \bibinfo{pages}{31--40}.
\newblock


\bibitem[Grinsztajn et~al\mbox{.}(2022)]%
        {grinsztajn2022tree}
\bibfield{author}{\bibinfo{person}{L{\'e}o Grinsztajn},
  \bibinfo{person}{Edouard Oyallon}, {and} \bibinfo{person}{Ga{\"e}l
  Varoquaux}.} \bibinfo{year}{2022}\natexlab{}.
\newblock \showarticletitle{Why do tree-based models still outperform deep
  learning on tabular data?}
\newblock \bibinfo{journal}{\emph{arXiv preprint arXiv:2207.08815}}
  (\bibinfo{year}{2022}).
\newblock


\bibitem[Hellerstein et~al\mbox{.}(2012)]%
        {hellerstein2012madlib}
\bibfield{author}{\bibinfo{person}{Joe Hellerstein},
  \bibinfo{person}{Christopher R{\'e}}, \bibinfo{person}{Florian Schoppmann},
  \bibinfo{person}{Daisy~Zhe Wang}, \bibinfo{person}{Eugene Fratkin},
  \bibinfo{person}{Aleksander Gorajek}, \bibinfo{person}{Kee~Siong Ng},
  \bibinfo{person}{Caleb Welton}, \bibinfo{person}{Xixuan Feng},
  \bibinfo{person}{Kun Li}, {et~al\mbox{.}}} \bibinfo{year}{2012}\natexlab{}.
\newblock \showarticletitle{The MADlib analytics library or MAD skills, the
  SQL}.
\newblock \bibinfo{journal}{\emph{arXiv preprint arXiv:1208.4165}}
  (\bibinfo{year}{2012}).
\newblock


\bibitem[Hu et~al\mbox{.}(2021)]%
        {hu2021tcudb}
\bibfield{author}{\bibinfo{person}{Yu-Ching Hu}, \bibinfo{person}{Yuliang Li},
  {and} \bibinfo{person}{Hung-Wei Tseng}.} \bibinfo{year}{2021}\natexlab{}.
\newblock \showarticletitle{TCUDB: Accelerating Database with Tensor
  Processors}.
\newblock \bibinfo{journal}{\emph{arXiv preprint arXiv:2112.07552}}
  (\bibinfo{year}{2021}).
\newblock


\bibitem[Huang et~al\mbox{.}(2023b)]%
        {joinboostgpu}
\bibfield{author}{\bibinfo{person}{Zezhou Huang}, \bibinfo{person}{Pavan~Kalyan
  Damalapati}, \bibinfo{person}{Rathijit Sen}, {and} \bibinfo{person}{Eugene
  Wu}.} \bibinfo{year}{2023}\natexlab{b}.
\newblock \showarticletitle{Random Forests over Normalized Data in CPU-GPU
  DBMSes}.
\newblock In \bibinfo{booktitle}{\emph{Data Management on New Hardware}}.
\newblock


\bibitem[Huang et~al\mbox{.}(2023a)]%
        {weighing}
\bibfield{author}{\bibinfo{person}{Zezhou Huang}, \bibinfo{person}{Pavan~Kalyan
  Damalapati}, {and} \bibinfo{person}{Eugene Wu}.}
  \bibinfo{year}{2023}\natexlab{a}.
\newblock \showarticletitle{Aggregation Consistency Errors in Semantic Layers
  and How to Avoid Them}. In \bibinfo{booktitle}{\emph{Proceedings of the
  Workshop on Human-In-the-Loop Data Analytics}}.
\newblock


\bibitem[Huang and Wu(2022)]%
        {cjt}
\bibfield{author}{\bibinfo{person}{Zezhou Huang} {and} \bibinfo{person}{Eugene
  Wu}.} \bibinfo{year}{2022}\natexlab{}.
\newblock \bibinfo{title}{Calibration: A Simple Trick for Wide-table Delta
  Analytics}.
\newblock
\newblock
\showeprint[arxiv]{2210.03851}~[cs.DB]


\bibitem[Huggins et~al\mbox{.}(2017)]%
        {huggins2017pass}
\bibfield{author}{\bibinfo{person}{Jonathan Huggins}, \bibinfo{person}{Ryan~P
  Adams}, {and} \bibinfo{person}{Tamara Broderick}.}
  \bibinfo{year}{2017}\natexlab{}.
\newblock \showarticletitle{Pass-glm: polynomial approximate sufficient
  statistics for scalable bayesian glm inference}.
\newblock \bibinfo{journal}{\emph{Advances in Neural Information Processing
  Systems}}  \bibinfo{volume}{30} (\bibinfo{year}{2017}).
\newblock


\bibitem[Jankov et~al\mbox{.}(2019)]%
        {jankov2019declarative}
\bibfield{author}{\bibinfo{person}{Dimitrije Jankov}, \bibinfo{person}{Shangyu
  Luo}, \bibinfo{person}{Binhang Yuan}, \bibinfo{person}{Zhuhua Cai},
  \bibinfo{person}{Jia Zou}, \bibinfo{person}{Chris Jermaine}, {and}
  \bibinfo{person}{Zekai~J Gao}.} \bibinfo{year}{2019}\natexlab{}.
\newblock \showarticletitle{Declarative recursive computation on an rdbms, or,
  why you should use a database for distributed machine learning}.
\newblock \bibinfo{journal}{\emph{arXiv preprint arXiv:1904.11121}}
  (\bibinfo{year}{2019}).
\newblock


\bibitem[Jankov et~al\mbox{.}(2021)]%
        {jankov2021distributed}
\bibfield{author}{\bibinfo{person}{Dimitrije Jankov}, \bibinfo{person}{Binhang
  Yuan}, \bibinfo{person}{Shangyu Luo}, {and} \bibinfo{person}{Chris
  Jermaine}.} \bibinfo{year}{2021}\natexlab{}.
\newblock \showarticletitle{Distributed numerical and machine learning
  computations via two-phase execution of aggregated join trees}.
\newblock \bibinfo{journal}{\emph{Proceedings of the VLDB Endowment}}
  \bibinfo{volume}{14}, \bibinfo{number}{7} (\bibinfo{year}{2021}).
\newblock


\bibitem[Joglekar et~al\mbox{.}(2015)]%
        {joglekar2015aggregations}
\bibfield{author}{\bibinfo{person}{Manas Joglekar}, \bibinfo{person}{Rohan
  Puttagunta}, {and} \bibinfo{person}{Christopher R{\'e}}.}
  \bibinfo{year}{2015}\natexlab{}.
\newblock \showarticletitle{Aggregations over generalized hypertree
  decompositions}.
\newblock \bibinfo{journal}{\emph{arXiv preprint arXiv:1508.07532}}
  (\bibinfo{year}{2015}).
\newblock


\bibitem[Joglekar et~al\mbox{.}(2016)]%
        {joglekar2016ajar}
\bibfield{author}{\bibinfo{person}{Manas~R Joglekar}, \bibinfo{person}{Rohan
  Puttagunta}, {and} \bibinfo{person}{Christopher R{\'e}}.}
  \bibinfo{year}{2016}\natexlab{}.
\newblock \showarticletitle{Ajar: Aggregations and joins over annotated
  relations}. In \bibinfo{booktitle}{\emph{Proceedings of the 35th ACM
  SIGMOD-SIGACT-SIGAI Symposium on Principles of Database Systems}}.
  \bibinfo{pages}{91--106}.
\newblock


\bibitem[Ke et~al\mbox{.}(2017)]%
        {ke2017lightgbm}
\bibfield{author}{\bibinfo{person}{Guolin Ke}, \bibinfo{person}{Qi Meng},
  \bibinfo{person}{Thomas Finley}, \bibinfo{person}{Taifeng Wang},
  \bibinfo{person}{Wei Chen}, \bibinfo{person}{Weidong Ma},
  \bibinfo{person}{Qiwei Ye}, {and} \bibinfo{person}{Tie-Yan Liu}.}
  \bibinfo{year}{2017}\natexlab{}.
\newblock \showarticletitle{Lightgbm: A highly efficient gradient boosting
  decision tree}.
\newblock \bibinfo{journal}{\emph{Advances in neural information processing
  systems}}  \bibinfo{volume}{30} (\bibinfo{year}{2017}).
\newblock


\bibitem[Khamis et~al\mbox{.}(2020)]%
        {khamis2020functional}
\bibfield{author}{\bibinfo{person}{Mahmoud~Abo Khamis}, \bibinfo{person}{Ryan~R
  Curtin}, \bibinfo{person}{Benjamin Moseley}, \bibinfo{person}{Hung~Q Ngo},
  \bibinfo{person}{XuanLong Nguyen}, \bibinfo{person}{Dan Olteanu}, {and}
  \bibinfo{person}{Maximilian Schleich}.} \bibinfo{year}{2020}\natexlab{}.
\newblock \showarticletitle{Functional Aggregate Queries with Additive
  Inequalities}.
\newblock \bibinfo{journal}{\emph{ACM Transactions on Database Systems (TODS)}}
  \bibinfo{volume}{45}, \bibinfo{number}{4} (\bibinfo{year}{2020}),
  \bibinfo{pages}{1--41}.
\newblock


\bibitem[Khamis et~al\mbox{.}(2018)]%
        {khamis2018ac}
\bibfield{author}{\bibinfo{person}{Mahmoud~Abo Khamis}, \bibinfo{person}{Hung~Q
  Ngo}, \bibinfo{person}{XuanLong Nguyen}, \bibinfo{person}{Dan Olteanu}, {and}
  \bibinfo{person}{Maximilian Schleich}.} \bibinfo{year}{2018}\natexlab{}.
\newblock \showarticletitle{AC/DC: In-database learning thunderstruck}. In
  \bibinfo{booktitle}{\emph{Proceedings of the Second Workshop on Data
  Management for End-To-End Machine Learning}}. \bibinfo{pages}{1--10}.
\newblock


\bibitem[Kobis(2017)]%
        {kobis2017learning}
\bibfield{author}{\bibinfo{person}{Lukas Kobis}.}
  \bibinfo{year}{2017}\natexlab{}.
\newblock \showarticletitle{Learning Decision Trees over Factorized Joins}.
\newblock  (\bibinfo{year}{2017}).
\newblock


\bibitem[Kraska et~al\mbox{.}(2013)]%
        {kraska2013mlbase}
\bibfield{author}{\bibinfo{person}{Tim Kraska}, \bibinfo{person}{Ameet
  Talwalkar}, \bibinfo{person}{John~C Duchi}, \bibinfo{person}{Rean Griffith},
  \bibinfo{person}{Michael~J Franklin}, {and} \bibinfo{person}{Michael~I
  Jordan}.} \bibinfo{year}{2013}\natexlab{}.
\newblock \showarticletitle{MLbase: A Distributed Machine-learning System.}. In
  \bibinfo{booktitle}{\emph{Cidr}}, Vol.~\bibinfo{volume}{1}.
  \bibinfo{pages}{2--1}.
\newblock


\bibitem[Kumar et~al\mbox{.}(2015)]%
        {kumar2015learning}
\bibfield{author}{\bibinfo{person}{Arun Kumar}, \bibinfo{person}{Jeffrey
  Naughton}, {and} \bibinfo{person}{Jignesh~M Patel}.}
  \bibinfo{year}{2015}\natexlab{}.
\newblock \showarticletitle{Learning generalized linear models over normalized
  data}. In \bibinfo{booktitle}{\emph{Proceedings of the 2015 ACM SIGMOD
  International Conference on Management of Data}}.
  \bibinfo{pages}{1969--1984}.
\newblock


\bibitem[Li et~al\mbox{.}(2017)]%
        {li2017mlog}
\bibfield{author}{\bibinfo{person}{Xupeng Li}, \bibinfo{person}{Bin Cui},
  \bibinfo{person}{Yiru Chen}, \bibinfo{person}{Wentao Wu}, {and}
  \bibinfo{person}{Ce Zhang}.} \bibinfo{year}{2017}\natexlab{}.
\newblock \showarticletitle{Mlog: Towards declarative in-database machine
  learning}.
\newblock \bibinfo{journal}{\emph{Proceedings of the VLDB Endowment}}
  \bibinfo{volume}{10}, \bibinfo{number}{12} (\bibinfo{year}{2017}),
  \bibinfo{pages}{1933--1936}.
\newblock


\bibitem[Mucchetti(2020)]%
        {mucchetti2020bigquery}
\bibfield{author}{\bibinfo{person}{Mark Mucchetti}.}
  \bibinfo{year}{2020}\natexlab{}.
\newblock \showarticletitle{BigQuery ML}.
\newblock In \bibinfo{booktitle}{\emph{BigQuery for Data Warehousing}}.
  \bibinfo{publisher}{Springer}, \bibinfo{pages}{419--468}.
\newblock


\bibitem[Murphy(2012)]%
        {murphy2012machine}
\bibfield{author}{\bibinfo{person}{Kevin~P Murphy}.}
  \bibinfo{year}{2012}\natexlab{}.
\newblock \bibinfo{booktitle}{\emph{Machine learning: a probabilistic
  perspective}}.
\newblock \bibinfo{publisher}{MIT press}.
\newblock


\bibitem[Neumann et~al\mbox{.}(2015)]%
        {neumann2015fast}
\bibfield{author}{\bibinfo{person}{Thomas Neumann}, \bibinfo{person}{Tobias
  M{\"u}hlbauer}, {and} \bibinfo{person}{Alfons Kemper}.}
  \bibinfo{year}{2015}\natexlab{}.
\newblock \showarticletitle{Fast serializable multi-version concurrency control
  for main-memory database systems}. In \bibinfo{booktitle}{\emph{Proceedings
  of the 2015 ACM SIGMOD International Conference on Management of Data}}.
  \bibinfo{pages}{677--689}.
\newblock


\bibitem[Nikolic and Olteanu(2018)]%
        {nikolic2018incremental}
\bibfield{author}{\bibinfo{person}{Milos Nikolic} {and} \bibinfo{person}{Dan
  Olteanu}.} \bibinfo{year}{2018}\natexlab{}.
\newblock \showarticletitle{Incremental view maintenance with triple lock
  factorization benefits}. In \bibinfo{booktitle}{\emph{Proceedings of the 2018
  International Conference on Management of Data}}. \bibinfo{pages}{365--380}.
\newblock


\bibitem[Olteanu and Schleich(2016)]%
        {olteanu2016factorized}
\bibfield{author}{\bibinfo{person}{Dan Olteanu} {and}
  \bibinfo{person}{Maximilian Schleich}.} \bibinfo{year}{2016}\natexlab{}.
\newblock \showarticletitle{Factorized databases}.
\newblock \bibinfo{journal}{\emph{ACM SIGMOD Record}} \bibinfo{volume}{45},
  \bibinfo{number}{2} (\bibinfo{year}{2016}), \bibinfo{pages}{5--16}.
\newblock


\bibitem[Olteanu and Z{\'a}vodn{\`y}(2015)]%
        {olteanu2015size}
\bibfield{author}{\bibinfo{person}{Dan Olteanu} {and} \bibinfo{person}{Jakub
  Z{\'a}vodn{\`y}}.} \bibinfo{year}{2015}\natexlab{}.
\newblock \showarticletitle{Size bounds for factorised representations of query
  results}.
\newblock \bibinfo{journal}{\emph{ACM Transactions on Database Systems (TODS)}}
  \bibinfo{volume}{40}, \bibinfo{number}{1} (\bibinfo{year}{2015}),
  \bibinfo{pages}{1--44}.
\newblock


\bibitem[pandas~development team(2020)]%
        {reback2020pandas}
\bibfield{author}{\bibinfo{person}{The pandas~development team}.}
  \bibinfo{year}{2020}\natexlab{}.
\newblock \bibinfo{booktitle}{\emph{pandas-dev/pandas: Pandas}}.
\newblock
\urldef\tempurl%
\url{https://doi.org/10.5281/zenodo.3509134}
\showDOI{\tempurl}


\bibitem[Paul et~al\mbox{.}(2021)]%
        {paul2021database}
\bibfield{author}{\bibinfo{person}{Johns Paul}, \bibinfo{person}{Shengliang
  Lu}, \bibinfo{person}{Bingsheng He}, {et~al\mbox{.}}}
  \bibinfo{year}{2021}\natexlab{}.
\newblock \showarticletitle{Database Systems on GPUs}.
\newblock \bibinfo{journal}{\emph{Foundations and Trends{\textregistered} in
  Databases}} \bibinfo{volume}{11}, \bibinfo{number}{1} (\bibinfo{year}{2021}),
  \bibinfo{pages}{1--108}.
\newblock


\bibitem[Pearl(1982)]%
        {pearl1982reverend}
\bibfield{author}{\bibinfo{person}{Judea Pearl}.}
  \bibinfo{year}{1982}\natexlab{}.
\newblock \bibinfo{booktitle}{\emph{Reverend Bayes on inference engines: A
  distributed hierarchical approach}}.
\newblock \bibinfo{publisher}{Cognitive Systems Laboratory, School of
  Engineering and Applied Science~…}.
\newblock


\bibitem[Pedregosa et~al\mbox{.}(2011)]%
        {scikit-learn}
\bibfield{author}{\bibinfo{person}{F. Pedregosa}, \bibinfo{person}{G.
  Varoquaux}, \bibinfo{person}{A. Gramfort}, \bibinfo{person}{V. Michel},
  \bibinfo{person}{B. Thirion}, \bibinfo{person}{O. Grisel},
  \bibinfo{person}{M. Blondel}, \bibinfo{person}{P. Prettenhofer},
  \bibinfo{person}{R. Weiss}, \bibinfo{person}{V. Dubourg}, \bibinfo{person}{J.
  Vanderplas}, \bibinfo{person}{A. Passos}, \bibinfo{person}{D. Cournapeau},
  \bibinfo{person}{M. Brucher}, \bibinfo{person}{M. Perrot}, {and}
  \bibinfo{person}{E. Duchesnay}.} \bibinfo{year}{2011}\natexlab{}.
\newblock \showarticletitle{Scikit-learn: Machine Learning in {P}ython}.
\newblock \bibinfo{journal}{\emph{Journal of Machine Learning Research}}
  \bibinfo{volume}{12} (\bibinfo{year}{2011}), \bibinfo{pages}{2825--2830}.
\newblock


\bibitem[Raasveldt and M{\"u}hleisen(2019)]%
        {raasveldt2019duckdb}
\bibfield{author}{\bibinfo{person}{Mark Raasveldt} {and}
  \bibinfo{person}{Hannes M{\"u}hleisen}.} \bibinfo{year}{2019}\natexlab{}.
\newblock \showarticletitle{DuckDB: an embeddable analytical database}. In
  \bibinfo{booktitle}{\emph{Proceedings of the 2019 International Conference on
  Management of Data}}. \bibinfo{pages}{1981--1984}.
\newblock


\bibitem[Raman et~al\mbox{.}(2013)]%
        {raman2013db2}
\bibfield{author}{\bibinfo{person}{Vijayshankar Raman}, \bibinfo{person}{Gopi
  Attaluri}, \bibinfo{person}{Ronald Barber}, \bibinfo{person}{Naresh
  Chainani}, \bibinfo{person}{David Kalmuk}, \bibinfo{person}{Vincent
  KulandaiSamy}, \bibinfo{person}{Jens Leenstra}, \bibinfo{person}{Sam
  Lightstone}, \bibinfo{person}{Shaorong Liu}, \bibinfo{person}{Guy~M Lohman},
  {et~al\mbox{.}}} \bibinfo{year}{2013}\natexlab{}.
\newblock \showarticletitle{DB2 with BLU acceleration: So much more than just a
  column store}.
\newblock \bibinfo{journal}{\emph{Proceedings of the VLDB Endowment}}
  \bibinfo{volume}{6}, \bibinfo{number}{11} (\bibinfo{year}{2013}),
  \bibinfo{pages}{1080--1091}.
\newblock


\bibitem[Rocklin(2015)]%
        {rocklin2015dask}
\bibfield{author}{\bibinfo{person}{Matthew Rocklin}.}
  \bibinfo{year}{2015}\natexlab{}.
\newblock \showarticletitle{Dask: Parallel computation with blocked algorithms
  and task scheduling}. In \bibinfo{booktitle}{\emph{Proceedings of the 14th
  python in science conference}}, Vol.~\bibinfo{volume}{130}. Citeseer,
  \bibinfo{pages}{136}.
\newblock


\bibitem[Saxena and Agarwal(2014)]%
        {saxena2014data}
\bibfield{author}{\bibinfo{person}{Geetika Saxena} {and}
  \bibinfo{person}{Bharat~Bhushan Agarwal}.} \bibinfo{year}{2014}\natexlab{}.
\newblock \showarticletitle{Data Warehouse Designing: Dimensional Modelling and
  ER Modelling}.
\newblock \bibinfo{journal}{\emph{International Journal of Engineering
  Inventions}} \bibinfo{volume}{3}, \bibinfo{number}{9} (\bibinfo{year}{2014}),
  \bibinfo{pages}{28--34}.
\newblock


\bibitem[Schleich et~al\mbox{.}(2019)]%
        {schleich2019layered}
\bibfield{author}{\bibinfo{person}{Maximilian Schleich}, \bibinfo{person}{Dan
  Olteanu}, \bibinfo{person}{Mahmoud Abo~Khamis}, \bibinfo{person}{Hung~Q Ngo},
  {and} \bibinfo{person}{XuanLong Nguyen}.} \bibinfo{year}{2019}\natexlab{}.
\newblock \showarticletitle{A layered aggregate engine for analytics
  workloads}. In \bibinfo{booktitle}{\emph{Proceedings of the 2019
  International Conference on Management of Data}}.
  \bibinfo{pages}{1642--1659}.
\newblock


\bibitem[Schleich et~al\mbox{.}(2016)]%
        {schleich2016learning}
\bibfield{author}{\bibinfo{person}{Maximilian Schleich}, \bibinfo{person}{Dan
  Olteanu}, {and} \bibinfo{person}{Radu Ciucanu}.}
  \bibinfo{year}{2016}\natexlab{}.
\newblock \showarticletitle{Learning linear regression models over factorized
  joins}. In \bibinfo{booktitle}{\emph{Proceedings of the 2016 International
  Conference on Management of Data}}. \bibinfo{pages}{3--18}.
\newblock


\bibitem[Shanghooshabad et~al\mbox{.}(2021)]%
        {shanghooshabad2021pgmjoins}
\bibfield{author}{\bibinfo{person}{Ali~Mohammadi Shanghooshabad},
  \bibinfo{person}{Meghdad Kurmanji}, \bibinfo{person}{Qingzhi Ma},
  \bibinfo{person}{Michael Shekelyan}, \bibinfo{person}{Mehrdad Almasi}, {and}
  \bibinfo{person}{Peter Triantafillou}.} \bibinfo{year}{2021}\natexlab{}.
\newblock \showarticletitle{PGMJoins: Random Join Sampling with Graphical
  Models}. In \bibinfo{booktitle}{\emph{Proceedings of the 2021 International
  Conference on Management of Data}}. \bibinfo{pages}{1610--1622}.
\newblock


\bibitem[Shankar et~al\mbox{.}(2022)]%
        {shankar2022operationalizing}
\bibfield{author}{\bibinfo{person}{Shreya Shankar}, \bibinfo{person}{Rolando
  Garcia}, \bibinfo{person}{Joseph~M Hellerstein}, {and}
  \bibinfo{person}{Aditya~G Parameswaran}.} \bibinfo{year}{2022}\natexlab{}.
\newblock \showarticletitle{Operationalizing Machine Learning: An Interview
  Study}.
\newblock \bibinfo{journal}{\emph{arXiv preprint arXiv:2209.09125}}
  (\bibinfo{year}{2022}).
\newblock


\bibitem[Shi(2007)]%
        {shi2007best}
\bibfield{author}{\bibinfo{person}{Haijian Shi}.}
  \bibinfo{year}{2007}\natexlab{}.
\newblock \emph{\bibinfo{title}{Best-first decision tree learning}}.
\newblock \bibinfo{thesistype}{Ph.\,D. Dissertation}. \bibinfo{school}{The
  University of Waikato}.
\newblock


\bibitem[Smelcer(1995)]%
        {smelcer1995user}
\bibfield{author}{\bibinfo{person}{John~B Smelcer}.}
  \bibinfo{year}{1995}\natexlab{}.
\newblock \showarticletitle{User errors in database query composition}.
\newblock \bibinfo{journal}{\emph{International Journal of Human-Computer
  Studies}} \bibinfo{volume}{42}, \bibinfo{number}{4} (\bibinfo{year}{1995}),
  \bibinfo{pages}{353--381}.
\newblock


\bibitem[Stonebraker et~al\mbox{.}(2018)]%
        {stonebraker2018c}
\bibfield{author}{\bibinfo{person}{Mike Stonebraker}, \bibinfo{person}{Daniel~J
  Abadi}, \bibinfo{person}{Adam Batkin}, \bibinfo{person}{Xuedong Chen},
  \bibinfo{person}{Mitch Cherniack}, \bibinfo{person}{Miguel Ferreira},
  \bibinfo{person}{Edmond Lau}, \bibinfo{person}{Amerson Lin},
  \bibinfo{person}{Sam Madden}, \bibinfo{person}{Elizabeth O'Neil},
  {et~al\mbox{.}}} \bibinfo{year}{2018}\natexlab{}.
\newblock \showarticletitle{C-store: a column-oriented DBMS}.
\newblock In \bibinfo{booktitle}{\emph{Making Databases Work: the Pragmatic
  Wisdom of Michael Stonebraker}}. \bibinfo{pages}{491--518}.
\newblock


\bibitem[Stonebraker et~al\mbox{.}(2011)]%
        {stonebraker2011architecture}
\bibfield{author}{\bibinfo{person}{Michael Stonebraker}, \bibinfo{person}{Paul
  Brown}, \bibinfo{person}{Alex Poliakov}, {and} \bibinfo{person}{Suchi
  Raman}.} \bibinfo{year}{2011}\natexlab{}.
\newblock \showarticletitle{The architecture of SciDB}. In
  \bibinfo{booktitle}{\emph{Scientific and Statistical Database Management:
  23rd International Conference, SSDBM 2011, Portland, OR, USA, July 20-22,
  2011. Proceedings 23}}. Springer, \bibinfo{pages}{1--16}.
\newblock


\bibitem[Viswanath and Ramaswamy(1998)]%
        {viswanath1998join}
\bibfield{author}{\bibinfo{person}{Swarup Acharya Phillip B~Gibbons Viswanath}
  {and} \bibinfo{person}{Poosala~Sridhar Ramaswamy}.}
  \bibinfo{year}{1998}\natexlab{}.
\newblock \showarticletitle{Join Synopses for Approximate Query Answering}.
\newblock  (\bibinfo{year}{1998}).
\newblock


\bibitem[Xin et~al\mbox{.}(2013)]%
        {xin2013shark}
\bibfield{author}{\bibinfo{person}{Reynold~S Xin}, \bibinfo{person}{Josh
  Rosen}, \bibinfo{person}{Matei Zaharia}, \bibinfo{person}{Michael~J
  Franklin}, \bibinfo{person}{Scott Shenker}, {and} \bibinfo{person}{Ion
  Stoica}.} \bibinfo{year}{2013}\natexlab{}.
\newblock \showarticletitle{Shark: SQL and rich analytics at scale}. In
  \bibinfo{booktitle}{\emph{Proceedings of the 2013 ACM SIGMOD International
  Conference on Management of data}}. \bibinfo{pages}{13--24}.
\newblock


\bibitem[Yang et~al\mbox{.}(2020)]%
        {yang2020towards}
\bibfield{author}{\bibinfo{person}{Keyu Yang}, \bibinfo{person}{Yunjun Gao},
  \bibinfo{person}{Lei Liang}, \bibinfo{person}{Bin Yao},
  \bibinfo{person}{Shiting Wen}, {and} \bibinfo{person}{Gang Chen}.}
  \bibinfo{year}{2020}\natexlab{}.
\newblock \showarticletitle{Towards factorized svm with gaussian kernels over
  normalized data}. In \bibinfo{booktitle}{\emph{2020 IEEE 36th International
  Conference on Data Engineering (ICDE)}}. IEEE, \bibinfo{pages}{1453--1464}.
\newblock


\bibitem[Zhao et~al\mbox{.}(2018)]%
        {zhao2018random}
\bibfield{author}{\bibinfo{person}{Zhuoyue Zhao}, \bibinfo{person}{Robert
  Christensen}, \bibinfo{person}{Feifei Li}, \bibinfo{person}{Xiao Hu}, {and}
  \bibinfo{person}{Ke Yi}.} \bibinfo{year}{2018}\natexlab{}.
\newblock \showarticletitle{Random sampling over joins revisited}. In
  \bibinfo{booktitle}{\emph{Proceedings of the 2018 International Conference on
  Management of Data}}. \bibinfo{pages}{1525--1539}.
\newblock


\end{thebibliography}

\clearpage

\appendix

\section{Decision Tree From Semi-ring}
\label{sec:dt_semiring}

In this section, we present the algorithms to express the criteria of reduction in variance for regression, information gain, gini impurity and chi-square for classifications based on the semi-ring in \Cref{table:semiring}. Computing these criteria is at the heart of decision tree training. Our algorithm here is based on \texttt{Sklearn}\footnote{\url{https://scikit-learn.org/stable/modules/tree.html\#tree-mathematical-formulation}}.

\stitle{Aggregated Semi-ring}: It has been shown that~\cite{schleich2019layered,abo2016faq}, for variance semi-ring, the aggregated semi-ring $\gamma(R_\Join)$ is a 3-tuple $(C,S,Q)$ that represents the count $C=\sum_{t\in R_\Join}1$, sum of target variable $S=\sum_{t\in R_\Join}t[Y]$, and the sum of squares of target variable $Q=\sum_{t\in R_\Join}t[Y]^2$. For classification with k classes, the aggregated semi-ring $\gamma(R_\Join)$ is a $(k+1)$-tuple $(C,C^1,...,C^k)$ that represents the total count $C=\sum_{t\in R_\Join}1$ and count of each class $C^i=\sum_{t\in R_\Join}\mathbf{1}_{t[Y] = i}$.

\stitle{Reduction in variance for regression}: 
In the scope of variance semi-ring and given $R_{\Join} = R_1 \bowtie R_2 \dots \bowtie R_n$ and schema $S_1, \dots, S_n$, let the average of $R_{\bowtie}[y]$ be $\hat{y}$. The total variance of the target variable in $R_{\bowtie}$ can be computed as 
\begin{align*}
var(R_\Join) =& \sum_{t \in R_{\bowtie}}(t[y] - \hat{y})^2 \\
    =& \sum_{t \in R_{\bowtie}} t[y]^2 - 2 \sum_{t \in R_{\bowtie}} t[y] \cdot \hat{y} + \sum_{t \in R_{\bowtie}} \hat{y}^2 \\
    =& \sum_{t \in R_{\bowtie}} t[y]^2 - \sum_{t \in R_{\bowtie}} \hat{y}^2 = Q - S^2/C 
\end{align*}

The third equation holds because $\sum_{t \in R_{\bowtie}} t[y] = \sum_{t \in R_{\bowtie}} \hat{y}$. Consider a selection predicate $\sigma$ where the attributes in $\sigma$, which splits $R_{\bowtie}$ into $\sigma(R_{\bowtie})$ and $\bar{\sigma}(R_{\bowtie})$, and let the average of $y$ in $\sigma(R_{\bowtie})$ and $\bar{\sigma}(R_{\bowtie})$ be $\hat{y}_{\sigma}$ and $\hat{y}_{\bar{\sigma}}$, respectively. Let their aggregated semi-ring be $\gamma (\sigma(R_{\bowtie})) = (C_\sigma,S_\sigma,Q_\sigma)$ and $\gamma (\bar\sigma(R_{\bowtie})) = (C - C_\sigma,S - S_\sigma,Q - Q_\sigma)$.
Thus, the new variance of $\sigma(R_{\bowtie})$ and $\bar{\sigma}(R_{\bowtie})$ can be similarly computed as
\begin{align*}
    var(\sigma(R_\Join)) =& \sum_{t \in \sigma(R_{\bowtie})} t[y]^2 - \sum_{t \in \sigma(R_{\bowtie})} \hat{y}_{\sigma}^2 = Q_\sigma - S_\sigma^2 / C_\sigma \\
    var(\Bar\sigma(R_\Join)) =& \sum_{t \in \bar{\sigma}(R_{\bowtie})} t[y]^2 - \sum_{t \in \bar{\sigma}(R_{\bowtie})} \hat{y}_{\bar{\sigma}}^2 \\ =& (Q - Q_\sigma) - (S - S_\sigma)^2/(C-C_\sigma)
\end{align*}

Thus, the reduction in variance can be expressed as computations using variance semi-ring
\begin{align*}
    var(R_\Join) - (var(\sigma(R_\Join)) + var(\Bar\sigma(R_\Join))) \\= - S^2/C + S_\sigma^2 / C_\sigma + (S - S_\sigma)^2/(C-C_\sigma)
\end{align*}

As an optimization, we note Q term is canceled out for the reduction in variance. Therefore, we don't include Q during training.

To parse it into SQL for each feature A, we can compute its best split in one query using the window function.  Note that, to avoid overflow, we need to compute $s_{t}^2/c_{t}$ as $(s_{t}/c_{t})\times s_{t}$.

{\footnotesize
\begin{verbatim}
SELECT A,-(stotal/ctotal)*stotal+(s/c)*s+(stotal-s)/(ctotal-c)*(stotal-s)
FROM (SELECT A, SUM(c) OVER(ORDER BY A) as c, SUM(s) OVER(ORDER BY A) as s
    FROM (  SELECT A, s, c
            FROM R
            GROUP BY A
         ))
ORDER BY criteria DESC
LIMIT 1;
\end{verbatim}
}

\stitle{Classification}: 
In the scope of class count semi-ring and given $R_{\Join}$, we will show that the criteria for classification (information gain, Gini impurity, Chi-square) can be computed given the aggregated class count semi-ring structures. Then it follows the same logic that we do not have to materialize the full $R_{\bowtie}$. Classification criteria are based on the probability of each class, which could be computed as $p^{k} = C^k/C$. The different criteria can be computed as:

\begin{align*}
entropy(R_\Join) &= -\sum_{k=1}^K p^{k}\log p^{k} =  -\sum_{k=1}^K (C^k/C)log(C^k/C)\\
gini(R_\Join) &= 1 - \sum_{i=1}^K (p^{k})^2 = 1 - \sum_{i=1}^K (C^k/C)^2
\end{align*}

 Consider a selection predicate $\sigma$ where the attributes in $\sigma$, which splits $R_{\bowtie}$ into $\sigma(R_{\bowtie})$ and $\bar{\sigma}(R_{\bowtie})$. Let their aggregated semi-ring be $\gamma (\sigma(R_{\bowtie})) = (C_\sigma,C^1_\sigma,...,C^k_\sigma)$ and $\gamma (\bar\sigma(R_{\bowtie})) = (C_{\bar\sigma}, C^1_{\bar\sigma}, ..., C^k_{\bar\sigma}) = (C - C_{\sigma},C - C^1_{\sigma}, ..., C - C^k_{\sigma})$. We can compute the reduction after the split in a way similar to regression.

Chi-square of the split is computed as: 
\begin{align*}
    \chi^2 &= \sum_{i = 1}^K \left(\frac{(C_\sigma^i - C^iC_\sigma / C)^2}{C^iC_\sigma /C} + \frac{(C_{\bar{\sigma}}^i - C^iC_{\bar{\sigma}} / C)^2}{C^iC_{\bar{\sigma}} / C}\right)
\end{align*}

\section{Boosted Trees From Semi-ring}
\label{sec:semiring}

In this section, we discuss the details of building Gradient Boosting from Semi-ring. We use the semi-rings as defined in \Cref{table:semiring_extended}. 

\begin{table*}
\begin{center}
\setlength{\tabcolsep}{0.4em} 
\begin{tabular}{c c c c} 
  \textbf{Semi-ring} & \textbf{Zero/One} & \textbf{Operator} & \textbf{Lift} \\ 
  \hline
  \thd{
Regression \\
$(\mathbf{R},\mathbf{R})$
} & 
\thd{{\bf 0:} $(0,0)$ \\
{\bf 1:} $(1,0)$ }
&
\thd{
$(h_1, g_1) + (h_2, g_2) = (h_1 + h_2, g_1 + g_2)$\\
$(h_1, g_1) \times (h_2, g_2) = (h_1 h_2, g_1 h_2 + g_2 h_1)$
}
 & 
$(h(t),g(t))$\\

\hline
\thd{
Classification\\
$((\mathbf{R}, \mathbf{R}), ...,(\mathbf{R}, \mathbf{R}))$
}
&
\thd{
{\bf 0:} $((0,0), ...,(0,0))$ \\
{\bf 1:} $((1,0), ...,(1,0))$ 
}
& 
\thd{$((h^1_1, g^1_1), ..., (h^k_1, g^k_1)) + ((h^1_2, g^1_2), ..., (h^k_2, g^k_2))$\\
$= ((h^1_1 + h^1_2, g^1_1 + g^1_2), ...,(h^k_1 + h^k_2, g^k_1 + g^k_2)) $\\
$((h^1_1, g^1_1), ..., (h^k_1, g^k_1)) \times ((h^1_2, g^1_2), ..., (h^k_2, g^k_2))$\\
$= ((h^1_1 h^1_2, g^1_1 h^1_2 + g^1_2 h^1_1),...,(h^k_1 h^k_2, g^k_1 h^k_2 + g^k_2 h^k_1))$}
 & 
$((h^1(t),g^1(t)),...,(h^k(t),g^k(t)))$ \\
\hline
\end{tabular}
\end{center}
\caption{Gradient Semi-rings for gradient boosting.}
\label{table:semiring_extended}
\end{table*}

\begin{table*}
\small
\begin{center}
\setlength{\tabcolsep}{0.4em} 
\begin{tabular}{|c|c |c| c| c|} 
\hline \textbf{Task} &
  \textbf{Loss function}  & \textbf{Gradient $g(\cdot)$}  & \textbf{Hessian $h(\cdot)$}  & \textbf{Prediction $P$}\\ 
  \hline
 & L2/rmse: $(\epsilon)^2$& $\epsilon$ & 1 & $mean(\mathcal{E})$
\\\cline{2-5}
&L1/mae: $|\epsilon|$ & $sign(\epsilon)$ & 1  & $median(\mathcal{E})$ 
\\\cline{2-5}
& \thd{Huber Loss: 
$\begin{cases} 0.5 \epsilon ^2, |\epsilon| \leq \delta\\
\delta (|\epsilon| - 0.5\delta), else
\end{cases}$}  
& \thd{
$\begin{cases} \epsilon, |\epsilon| \leq \delta\\
\delta\cdot sign(\epsilon), else
\end{cases}$}  & 1 & $p^*$ 
 \\\cline{2-5}
&\thd{Fair Loss:
$c|\epsilon| - c^2log(|\epsilon|/c + 1)$}  
& $c\epsilon/(|\epsilon| + c)$  & $c^2/(|\epsilon| + c)^2$ &  $p^*$ 
 \\\cline{2-5}
& \thd{Poisson Loss: 
$e^p - yp$}  
 & $e^p - y$ &  $e^p$  &   $p^*$ 
 \\\cline{2-5}
Regression & \thd{Quantile Loss: 
$\begin{cases} (\alpha - 1)\epsilon, \epsilon < 0\\
\alpha \epsilon,  \epsilon \geq 0
\end{cases}$}  
& \thd{
$\begin{cases} \alpha - 1, \epsilon < 0\\
\alpha ,  \epsilon \geq 0
\end{cases}$}  & 1 & $ pctl_\alpha(\mathcal{E})$
 \\\cline{2-5}
 & \thd{mape: 
$|y-p|/max(1,|y|)$}  
& $\frac{sign(y-p)}{max(1,|y|)}$ 
& 1  & $median(\mathcal{E})$ 
 \\\cline{2-5}
&  \thd{Gamma Loss: 
$\frac{\frac{y}{p}- log(\frac{y}{p\psi})}{\psi}  + log(y) $}  
&  $1-ye^{-p}$ &  $ye^{-p}$ &   $p^*$ 
 \\\cline{2-5}
 &  \thd{Tweedie Loss: 
$-\frac{y e^{(1-\rho)log(p)}}{1-\rho} +$ \\
$\frac{e^{(2-\rho)log(p)}}{2-\rho}$}  
& \thd{$-ye^{(1-\rho)p} +$\\$e^{(2-\rho)p}$} &   \thd{$-(1-\rho)ye^{(1-\rho)p} +$\\$(2-\rho)e^{(2-\rho)p}$} &  $p^*$ 
 \\\hline
 \thd{
Classification}   & \thd{Softmax:
$-y^klog(p^k)$}  
& $p^k - y^k$ &  $\frac{K}{K-1} p^k(1-p^k)$ &   $p^*$ 
 \\\hline
\end{tabular}

\end{center}
\caption{Summary of Gradient and Hessian implemented in the source codes from \lgbm. We note that Gradients and Hessians are not mathematically rigorous; they have been normalized and approximated for practical concerns. The residual $\epsilon = y - p$. $\mathcal{E}$ is the residual column. $pctl_\alpha$ is $\alpha$ percentile function. }
\label{table:loss}
\end{table*}

\subsection{Semi-ring Extension}

\stitle{Extension to bag semantics and weighted relations.}  Tuples in relations could be weighted. For relation $R_i$, let $w$ be the function that maps tuple $t \in R_i$ to the weight as a real number. We modify the definition of semi-ring such that their count is a real number, then annotate relations as before, but their annotations are further multiplied by $(w(t),0,0)$ for weighting.

\stitle{Extension to theta joins and outer joins.} The support for theta joins over annotated relations is straightforward: 
$$(R\Join_\theta T)(t) = \begin{cases} 
R(\pi_{S_R} (t)) \times T(\pi_{S_T} (t)), \theta(t)\\
Zero \HS Element, \neg \theta(t)
\end{cases} 
$$

To support outer-join, note that the semantic of annotated relation is that tuples not in the relation are annotated with zero-element~\cite{abo2016faq}. This is undesirable as zero-element annihilates other elements when multiplied. Therefore for outer-join, we define the non-existed tuples to be annotated with one-element. Let $J = S_R \cap S_T$ be the join key between R and S:

$$
(R \leftouterjoin T)(t) = \begin{cases} 
R(\pi_{S_R} (t)) \times T(\pi_{S_T} (t)), \pi_{S_T} (t) \in T\\
R(\pi_{S_R} (t)), \pi_J(t) \not\in \pi_J(T) \wedge \pi_{S_T - J}(t)\HS ALL\HS NULL 
\end{cases} 
$$

\subsection{Boosted Trees Preliminary}
We provide the background of Boosted Tree based on ~\cite{chen2016xgboost,murphy2012machine,ke2017lightgbm}.

\stitle{Objective.} We consider ML model $f$ that maps tuple $t \in R_\Join$ to the prediction. The prediction is a real number for regression, and a set of real numbers corresponding to the probabilities of classes for classification. We want to train a model $f$ that minimizes the following objective:

$$L(f) = \sum_{t\in R_\Join} l(f(t), t[Y]) + \Omega(f)$$

\noindent Here, $l$ is the loss function and $\Omega$ is the regularization function.

\stitle{Regularization.} For boosted tree of K iterations, $f$ uses K additive functions $f_1,...,f_K$ for predictions: 

$$f(t) = \sum_{k=1}^K f_k(t)$$

\noindent The regularization term prefers tree models with smaller number of leaves and variance. For each tree $f_i$, let $T_i$ be its number of leaves and $p_i$ be a vector of leaf predictions. Then:

$$\Omega(f) = \alpha \sum_{k=1}^K T_k + 0.5\beta \sum_{k=1}^K ||p_i||^2 $$

\stitle{Optimization.} Boosted Trees iteratively train decision trees to optimize the objective based on the predictions from previous trees. For $ith$ tree $f_k$, it is optimizing:

$$L^i(f_i) = \sum_{t\in R_\Join} l(\sum_{k=1}^{i-1} f_k(t) + f_i(t), t[Y]) + \Omega(f_i)$$

\noindent After applying second optimization and removing constant terms:

$$\tilde{L}^i(f_i) = \sum_{t\in R_\Join} (g_t f_i(t) + 0.5 h_t f_i^2(t)) + \Omega(f_i)$$

\noindent where $g_t$ and $h_t$ are the gradients and hessians on the loss $l(\sum_{k=1}^{i-1} f_k(t), t[Y])$ for tuple $t$ with respect to the predictions from previous trees $\sum_{k=1}^{i-1} f_k(t)$.

Next, we study the optimal prediction for leaf node. Consider a leaf node of decision tree $f_i$ with number of leaves $T_i$, selection predicate $\sigma$ and prediction $p$ as a variable. The optimization objective (after removing constant terms) for this leaf node is then:

$$\tilde{L}^i_\sigma(p) = \sum_{t\in \sigma(R_\Join)} (g_t p + 0.5 h_t p^2) + 0.5 \beta p^2$$

\noindent The optimal prediction is then:

$$p^* = - \frac{\sum_{t\in \sigma(R_\Join)} g_t }{\sum_{t\in \sigma(R_\Join)} h_t + \beta}$$

\noindent And the objective becomes:

$$\tilde{L}^i_\sigma(p^*) = -0.5 \frac{(\sum_{t\in \sigma(R_\Join)} g_t)^2}{\sum_{t\in \sigma(R_\Join)} h_t + \beta}$$

Finally, we study the problem of evaluating a split. Consider the selection predicate $\sigma$ that splits $R_\Join$ into $\sigma(R_\Join)$ and $\Bar\sigma(R_\Join)$. The initial loss before split is:

$$\tilde{L} = -0.5 \frac{(\sum_{t\in R_\Join} g_t)^2}{\sum_{t\in R_\Join} h_t + \beta}$$

\noindent The loss after split is:

$$\tilde{L}_\sigma + \tilde{L}_{\Bar{\sigma}} + \alpha = -0.5 [\frac{(\sum_{t\in \sigma(R_\Join)} g_t)^2}{\sum_{t\in \sigma(R_\Join)} h_t + \beta} + \frac{(\sum_{t\in \Bar\sigma(R_\Join)} g_t)^2}{\sum_{t\in \Bar\sigma(R_\Join)} h_t + \beta} ] + \alpha$$

\noindent The $\alpha$ term is because spliting increases the number of leaves. Finally, the reduction of loss from the split is:

$$\tilde{L} - (\tilde{L}_\sigma + \tilde{L}_{\Bar{\sigma}} + \alpha )$$

\noindent The training process of Boosted Trees evaluates the reduction of loss for all candidate splits and chooses the best one.

\subsection{Boosted Trees from Semi-ring}

From the last section, we find that the core statistics over relational tables are the sum of gradients and hessians on the loss. Therefore, we build gradient semi-ring and lift relation based on the  gradient and hessian for different objective functions, thus avoiding the materialization of $R_\Join$.

The definition of gradient semi-ring is in \Cref{table:semiring_extended} and the formula of gradients and hessians for different objective functions based on \lgbm are in \Cref{table:loss}. We note that the gradients and hessians are not mathematical rigourious, and some losses are approximated. For example, mean absolute error has hessian undefined when error = 0, and 0 for the rest. Hessian can't be zero, and thus the loss is approximated without second order gradient (so hessian all 1) ~\footnote{https://github.com/microsoft/LightGBM/pull/175}.

\section{Other Experiments}
\label{sec:tpcexp}
\begin{figure}

\begin{subfigure}[t]{0.45\textwidth}
     \centering
     \includegraphics[width=\textwidth]{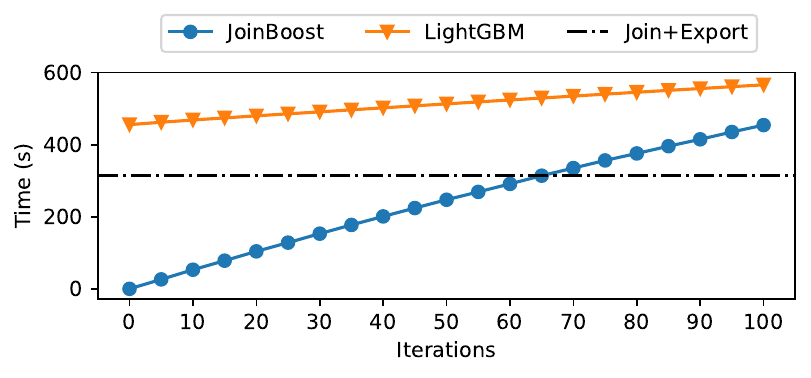}
      \caption{Gradient Boosting for TPC-DS.}
      \label{fig:tpcds_gb_time}
 \end{subfigure}
 \hfill
 \begin{subfigure}[t]{0.45\textwidth}
     \centering
     \includegraphics[width=\textwidth]{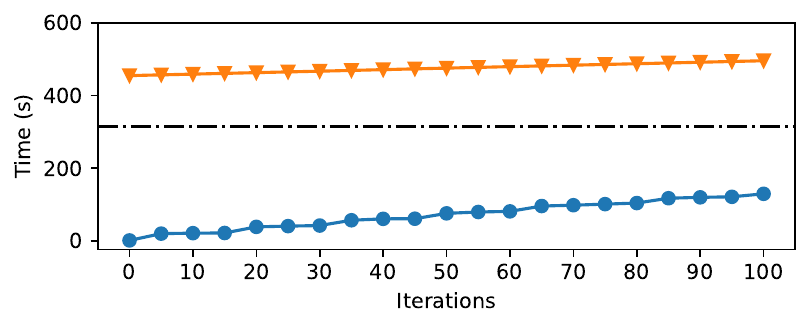}
      \caption{Random Forest for TPC-DS.}
      \label{fig:tpcds_rf_time}
 \end{subfigure}
 \hfill

\centering
  \begin{subfigure}[t]{0.45\textwidth}
     \centering
     \includegraphics[width=\textwidth]{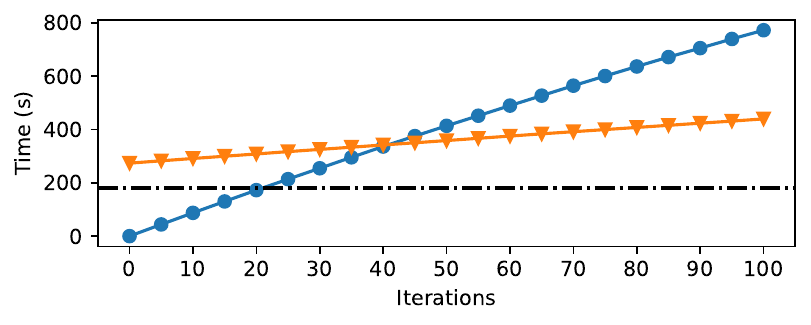}
      \caption{Gradient Boosting for TPC-H.}
      \label{fig:tpch_gb_time}
 \end{subfigure}
 \hfill
 \begin{subfigure}[t]{0.45\textwidth}
     \centering
     \includegraphics[width=\textwidth]{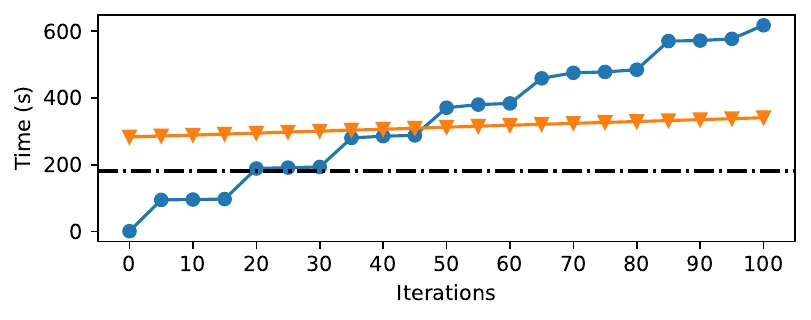}
      \caption{Random Forest for TPC-H.}
 \end{subfigure}
 \hfill

\caption{Gradient Boosting and Random Forest training time and model performance compared to the SOTA ML Frameworks. Dotted black line is the overhead of materializing the join and export join by DuckDB.}
\label{fig:tpc_exp}
\end{figure}

\subsection{Comparison with \lgbm}
\Cref{fig:tpc_exp} shows the experiment results for TPC-DS with SF=10. \sys is $\sim 3 \times$ faster than \lgbm for random forests, and  $\sim 1.3 \times$ faster than \lgbm for gradient boosting.
For TPC-H, \sys is slowed because of the large dimension tables (\texttt{Orders} and \texttt{PartSupp}). The messages between the fact table and these two dimension tables are large but expensive. To address this, the junction hypertree could be redesigned to reduce message sizes~\cite{cjt} as future  works.

\begin{figure}
\centering
\begin{subfigure}[b]{0.2\textwidth}
\includegraphics[width=\textwidth]{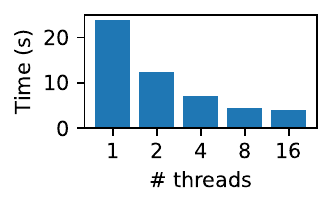}
\caption{}
\end{subfigure}
\hfill
\begin{subfigure}[b]{0.18\textwidth}
\includegraphics[width=\textwidth]{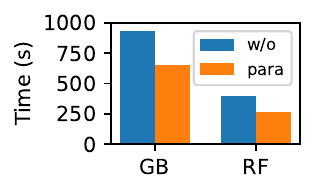}
\caption{}
\end{subfigure}
\hfill
\caption{(a) Training time of 1 tree of 8 leaves for intra-query parallelism with a varying number of threads.
(b) Training time of Gradient Boosting and Random Forest with (\texttt{para}) and without (\texttt{w/o}) inter-query parallelism. 
}
\label{fig:para_exp}
\end{figure}

\subsection{Parallelization}
\label{sec:para_exp}

We use \sys with (\texttt{para}) and without (\texttt{w/o}) inter-query parallelism on \texttt{Favorita}, but always with intra-query optimization (4 threads per query for \texttt{para} and 16 threads per query for \texttt{w/o}), to train a tree of 8 leaves. \Cref{fig:para_exp} shows the result.
Gradient Boosting and Random Forest with 100 trees \Cref{fig:para_exp} shows the model training time for 100 trees.  For gradient boosting, the training time is reduced by $28\%$ by exploiting partial dependency in queries. For Random Forest, the tree-wise parallelism reduces the training time more significantly by $35\%$.
\section{Other Optimizations}
\label{app:opt}

\begin{figure}
     \centering
  \includegraphics[width=0.8\columnwidth]{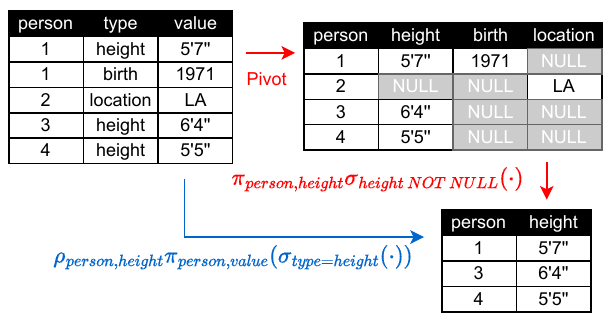}
  \vspace*{-3mm}
      \caption{
      Pivot Transformation Optimizations over $Person\_Info$ in IMDB. Naively, the sparse pivot result is materialized, then each feature (e.g., height) is selected to evaluate the split criteria (\red{red}). The query could be rewritten as a selection over type column by feature name, thus avoiding the materialization of a sparse pivot table (\blue{blue}).}
      \vspace*{-6mm}
      \label{fig:pivot}
\end{figure}

\subsection{Data Transformations.} 
Prior factorized learning work~\cite{khamis2018ac} introduced techniques that use functional dependencies to prune attributes prior to training, and to avoid  materializing one-hot encodings for categorical attributes.    A benefit of a middle-ware approach to in-DB ML is that we can leverage existing query optimizations to support other data transformations.  In addition to the prior techniques above, \sys introduces an optimization to support ``pivot'' transformations.   

ML algorithms traditionally expect a dense matrix representation of the features.  In contrast, DBMSes store sparse representations that need to be pivoted for ML libraries.  \Cref{fig:pivot} illustrates IMDB data containing attribute-value pairs (e.g., birth, 1971) that encode information about a person.  

If \texttt{type} were used as a feature, we first need to pivot the attribute into a sparse table with one column for each \texttt{type} value (e.g., height, birth).  This matrix is then aggregated to compute statistics during training.  To avoid this, Cunningham et al.~\cite{cunningham2004pivot} describe an optimization that rewrites aggregations over pivots as selections over the original table.   \sys easily supports such optimizations simply by adding an additional rewriting step. For the IMDB dataset, this optimization speeds up node splits by $3.8\times$ for the $Person\_Info$ table, compared to naive pivoting.

\subsection{Missing Join Keys.}
ML libraries like \lgbm and \xgb provide native support for missing values because they are ubiquitous in real-world datasets.  They evaluate losses by allocating missing values to either side of a candidate split and choosing the best one.   \sys efficiently supports this: candidate splits are already computed using a group-by, so the missing values will be aggregated in a group.    We simply add the missing value's annotation to each split candidate and choose the best.  The overhead is negligible compared to the aggregation itself.

What if the join key is missing values, as this affects the join's output cardinality?  Semi-ring annotations naturally support this: message passing simply uses full outer joins instead of inner joins, and then applies the procedure above.

\stitle{Identity and Semi-join Message Optimizations.} 
Snowflake schemas often exhibit a property that allows us to drop messages (and skip the associated joins) along {\it Identity Paths} during message passing.   
An identity path starts from a leaf relation (usually a dimension table), and flows along 1-to-N relationship edges, as long as each relation is not $R_Y$ and no join key is missing.   The key property is that semi-ring addition (aggregation) does not apply along the identity path due to the 1-to-N join relationships.

Let $L\to P$ connects leaf relation $L$ to $P$.  Each tuple $\in L$ is annotated with the 1 semi-ring element, and each tuple in $P$ joins with exactly one tuple in $L$.  
$m_{L\to P}$ is an {\it Identity Message} because it doesn't change $P$'s annotations, and thus can be dropped. Relations along the identity path are all initially annotated with $1$, so this applies recursively.
This optimization is used when splitting the root decision tree node.

Now suppose the best split was on attribute $l$ in $L$, so the edge is now $\sigma_{l}(L)\to P$.  Although $L$ does not emit an identity message, we know that each tuple in the message is annotated with $1$.  Thus, we don't materialize the semi-ring elements as columns and can rewrite the join with $P$ as a semi-join, and simply filter $P$ by the join values in $\sigma_{l}(L)$.

\begin{figure}
\centering
     \centering
     \includegraphics[width=0.47\textwidth]{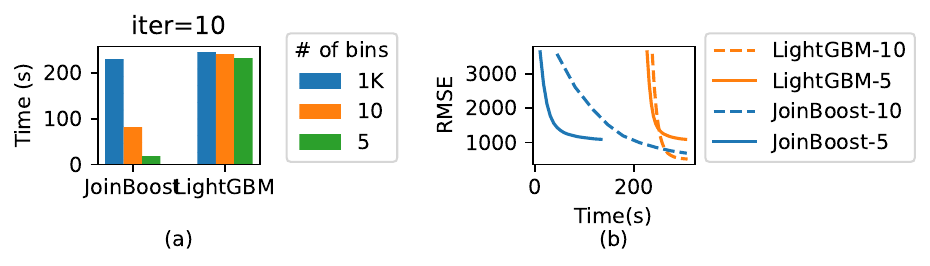}
     \vspace*{-5mm}
 \caption{Histogram-based Gradient Boosting. (a) \sys is much faster with fewer bins due to the smaller cuboid sizes. (b) Training time-accuracy trade-off. \sys with $\#bin=5$ pushes
the Pareto frontier and reaches $RMSE=1500$ in $20s$.}
\vspace*{-6mm}
 \label{fig:cubetime}
\end{figure}

\subsection{Histogram-based Cuboid.}\label{sec:cuboid}
A popular technique implemented in \lgbm and \xgb  is histogram-based gradient boosting. It computes histograms over each feature, and replaces each feature's value with its bin id (the cardinality is the same).  Varying the number of bins trades off accuracy for lower training time.   

A natural optimization is to leverage data cubes~\cite{gray1997data} when the number of bins is small and the data is sparse.  We evaluate a naive approach that materializes the full dimensional cuboid---\texttt{GROUP BY} all feature attributes---and uses it instead of factorized learning to execute the aggregation queries for training.  On the Favorita dataset, 5 bins result in a cuboid with $3M$ rows, as compared to 80M for the full join result.   Since training is based on semi-ring annotations, the rest of the algorithms remain the same.

For the experiment, we vary the number of histogram bins to study when using the cuboid optimization (\Cref{sec:cuboid}) is effective.  For bin=$5,10,1K$, the cuboid contains ${\sim}3M$, ${\sim}25M$, and ${\sim}80M$ rows respectively.  

\Cref{fig:cubetime}(a) shows that at iteration $10$ and $bin=5$, \sys speeds up by ${>}100\times$ the smaller cuboid size and faster aggregation queries, whereas \lgbm only slightly improves.  As we run more boosting iterations (\Cref{fig:cubetime}(b)) \sys-5 quickly converges (${\sim}20s$); \sys-10 takes longer to converge because the cuboid is larger but is still competitive to \lgbm.  In theory, we could gradually increase the number of bins throughout training to further improve model accuracy, but leave this to future work.

\end{document}